\documentclass[onecolumn,trackchanges]{aastex701}

\usepackage{hyperref}
\usepackage{amsmath}
\usepackage{amssymb}
\usepackage{xcolor}
\usepackage{comment}
\usepackage{makecell}
\usepackage{array}
\usepackage{multirow}
\usepackage{adjustbox,booktabs}

\begin{document}

\title{Revisiting the Multiwavelength Data of GRB 240825A: Implications for a Two-Stage Central Engine and a Dusty Environment}
\correspondingauthor{Xiao-Li Huang, En-Wei Liang}

\author[0009-0003-0580-6724]{Ren-Jie Xiong}
\affiliation{Guangxi Key Laboratory for Relativistic Astrophysics, School of Physical Science and Technology, Guangxi University, Nanning 530004, People's Republic of China}
\email{xiongrenj@st.gxu.edu.cn}

\author[0009-0001-8025-3205]{Zi-Qi Wang}
\affiliation{Guangxi Key Laboratory for Relativistic Astrophysics, School of Physical Science and Technology, Guangxi University, Nanning 530004, People's Republic of China}
\email{ziqi.wang@st.gxu.edu.cn}

\author[0000-0002-9725-7114]{Xiao-Li Huang}
\affiliation{School of Physics and Electronic Science, Guizhou Normal University, Guiyang 550025, People’s Republic of China}
\affiliation{National Astronomical Observatories, Chinese Academy of Sciences, Beijing 100101, People's Republic of China}
\email[show]{xiaoli.huang@gznu.edu.cn}

\author[0000-0002-7044-733X]{En-Wei Liang}
\affiliation{Guangxi Key Laboratory for Relativistic Astrophysics, School of Physical Science and Technology, Guangxi University, Nanning 530004, People's Republic of China}
\email[show]{lew@gxu.edu.cn}

\begin{abstract}
Reanalyzing the multi-wavelength data of GRB 240825A, we identify two distinct prompt-emission episodes, i.e. an early, spectrally soft plateau and a subsequent bright, spectrally hard gamma-ray pulse. We show that this signature is consistent with a two-stage central engine model, in which the early soft plateau is powered by the dipolar spin-down radiation of a transient proto-magnetar, while the later hard pulse arises from a relativistic jet launched by fallback accretion onto the nascent black hole via the Blandford–Znajek mechanism. Adopting simulation-motivated initial black hole (BH) masses ($M_{\rm BH,0}$) and spins ($a_0$), our fits to the second episode yield total masses supplied to the post-collapse disk of $M_{\rm fb}=0.45$--$0.96\,M_\odot$, fallback radii of $r_{\rm fb}=16.9$--$17.6\,r_g$, and characteristic fallback timescales of $t_{\rm fb}=0.13$--$0.25\,\mathrm{s}$. The corresponding maximum inner accretion rates range from $0.18$ to $0.70\,M_\odot\,\mathrm{s}^{-1}$, which is sensitive to $a_0$. Joint spectral fits with optical–X-ray afterglow data reveal a coupled temporal evolution of dust extinction and equivalent hydrogen column density. Incorporating this extinction evolution, an external forward–reverse shock model in a wind-like medium can account for the radio-to-GeV band afterglow data. Although ambient dust and the bright host galaxy may obscure any associated supernova, late-time optical reddening remains compatible with a supernova contribution. Applying the empirical relation between the prompt spectral peak energy and the SN peak bolometric magnitude gives an illustrative estimate of $M_{\rm bol,peak}=-19.46\pm0.35$~mag for a putative associated SN. These results support a two-stage evolution of the central engine and variable dusty environment of GRB~240825A, which provides insights into the nature of its progenitor.
\end{abstract}


\section{Introduction}
\label{Introduction}
Long gamma-ray bursts (LGRBs) are characterized by bright and highly variable gamma-ray emission in the keV--MeV band, with durations ($T_{90}$) longer than 2 s \citep{1993ApJ...413L.101K}. Their association with broad-lined Type Ic supernovae establishes a massive-star core-collapse origin \citep{1998Natur.395..670G,2003Natur.423..847H,2003ApJ...591L..17S,2006ARA&A..44..507W}. Compact Wolf--Rayet stars are widely regarded as the leading progenitor candidates for LGRBs \citep{1993ApJ...405..273W, 1999ApJ...524..262M, 2003ApJ...595..935M}. However, extended massive-star progenitors \citep{2013ApJ...766...30G, 2013ApJ...778...67N} and alternative binary evolutionary channels have also been proposed \citep{2004ApJ...607L..17P, 2005ApJ...623..302F}. These progenitor channels can produce different pre-collapse mass-loss histories, leaving distinct signatures in the circumburst environment \citep{2003ApJ...591..288H}. Stellar winds, eruptive mass-loss episodes, and binary interactions can reshape the circumburst medium, setting the density profile, composition, and inhomogeneity \citep{2000ApJ...536..195C, 2015ApJ...805..159M}. Multiwavelength afterglow observations therefore provide a powerful probe of the circumburst environment and constrain the late-stage evolution of LGRB progenitors \citep{1998ApJ...497L..17S, 2008ApJ...672..433S, 2010MNRAS.401.2773S, 2011A&A...534A.108K}.

The central engines of LGRBs are generally attributed to either hyper-accreting black holes or newly born magnetars \citep{1992Natur.357..472U, 1992ApJ...392L...9D, 1999ApJ...518..356P, 2011MNRAS.413.2031M}. Nevertheless, the central engine remains observationally elusive. The launch of the \textit{Swift} satellite enabled systematic early X-ray observations, revealing the canonical X-ray afterglow behavior and providing new insights into central-engine activity \citep{2006ApJ...642..389N,2006ApJ...642..354Z}. Some shallow-decay X-ray plateaus can be explained by sustained energy injection into the external shock, for example through magnetar spin-down \citep{1998A&A...333L..87D,2001ApJ...552L..35Z}. A distinct class of X-ray plateaus, known as internal plateaus, is followed by a very steep decay that is difficult to explain with standard external-shock emission \citep{2007ApJ...665..599T,2010MNRAS.402..705L}. In the magnetar interpretation, this abrupt decline may mark the collapse of a supramassive neutron star into a black hole, allowing the plateau duration to constrain its lifetime \citep{2014ApJ...785...74L}. High-angular-momentum stellar material of collapsars can circularize into a disk on timescales of seconds \citep{1999ApJ...524..262M, 2008ApJ...679..639Z, 2023ApJ...957...68B, 2024ApJ...976L..13G}. The renewed ultra-relativistic jet can then be powered by the neutrino-antineutrino annihilation or the Blandford-Znajek (BZ) mechanism. In the former, neutrinos and antineutrinos carry away part of the gravitational binding energy released as heat in the disk and annihilate above it to power a fireball \citep{1992MNRAS.257P..29M, 2010A&A...516A..16L}. The latter extracts the black hole's rotational energy through large-scale magnetic flux threading its horizon \citep{1977MNRAS.179..433B, 2000PhR...325...83L, 2017ApJ...849..119C}. 

For a given high black-hole spin and inner accretion rate, the BZ power can exceed the neutrino annihilation luminosity by roughly two orders of magnitude under the adopted magnetic-field assumptions \citep{2015ApJS..218...12L,2024MNRAS.527.7905L}. Neutrino annihilation is strongly suppressed when the inner accretion rate falls below the threshold for efficient neutrino cooling \citep{2011MNRAS.410.2302Z}. If the disc outflows are considered, the above conclusion will be strengthened. Nevertheless, neutrino annihilation may dominate in some systems \citep{2017NewAR..79....1L}. Here, we adopt the BZ mechanism to model the jet after black-hole formation, with viscous transport and winds in the fallback-fed disk regulating the inner accretion rate. However, a Blandford–Znajek-powered jet requires both rapid rotation and a strong magnetic field. Seed magnetic fields carried by the collapsing stellar material can be amplified in the disk by differential rotation and turbulence driven by the magnetorotational instability \citep{1991ApJ...376..214B,1995ApJ...440..742H}. Local field amplification alone, however, does not ensure the coherent, large-scale poloidal flux required to sustain a powerful BZ jet \citep{2024ApJ...976L..13G}. In the two-stage scenario considered here, the preceding magnetized proto-neutron star (PNS) provides a plausible source of this flux. Following the collapse of the PNS, the newly formed black hole may inherit magnetic flux from the star, with the surrounding accretion disk confining this flux near the horizon to support a BZ-powered jet \citep{2024ApJ...976L..13G}. This evolutionary sequence could underlie a transition in the prompt emission from magnetar spin-down to a BZ-powered jet sustained by fallback accretion. Direct observational evidence connecting these stages in an individual GRB, however, remains limited.

GRB~240825A ($z=0.659$; \citealt{2024GCN.37293....1M}) is a well-observed LGRB detected by \textit{Swift}/BAT, \textit{Fermi}/GBM--LAT, and Konus--Wind. It exhibits several collapsar-like properties, including $T_{90}\sim4$~s, a dusty star-forming host galaxy, and consistency with the Amati relation. In addition, its prompt emission features a distinctive GeV excess with a spectral break at $\sim50$~MeV \citep{2025ApJ...984L..45Z,2025ApJ...985L..30W}. Multiband optical observations reveal an unusually reddened afterglow requiring substantial host extinction ($E_{B-V}^{\rm host}\sim0.37$--$0.57$~mag), while the gradual color evolution toward bluer wavelengths suggests possible dust destruction by the burst radiation field \citep{2025RAA....25j5003W,2025ApJ...979...38C}. The consistent evolution of the optical-to-X-ray spectral index ($\beta_{\rm OX}$) indicates a transient suppression of optical emission followed by recovery \citep{2026ApJ...997..101L}. Together with the lack of a SN 1998bw-like bump, these observations indicate a complex progenitor and explosion scenario for GRB 240825A \citep{2025arXiv250800142G}. These observations make GRB~240825A an exceptional laboratory for exploring the interplay between prompt-emission evolution, central-engine activity, and the circumburst environment. 

In this work, we focus on the evolution of the central engine, the properties of the afterglow jet, and the dusty circumburst environment of GRB~240825A. This paper is organized as follows. In Section~\ref{sec:Prompt Gamma-ray Emission and Implications for its Central Engine Activity}, we analyze the temporal properties of the prompt emission within a two-step collapse central-engine framework. Section~\ref{sec:Properties of Afterglow Jet and its Environment} presents the multi-wavelength afterglow analysis and modeling, incorporating time-dependent dust extinction. In Section~\ref{sec:Late-time Optical Constraints on a Possible Supernova Component}, we discuss the possible presence of a weak, host-obscured supernova component associated with GRB~240825A. The results are discussed and summarized in Section~\ref{sect:Discussion and Conclusions}. Throughout this paper, we adopt the standard notation $Q_n = Q/10^n$ in cgs units and a flat $\Lambda$CDM cosmology with $H_0 = 67.4~\mathrm{km\,s^{-1}\,Mpc^{-1}}$, $\Omega_{\mathrm{m}} = 0.315$, and $\Omega_{\Lambda} = 0.685$ \citep{2020A&A...641A...6P}.

\section{Prompt Gamma-ray Emission and Implications for its Central Engine Activity}
\label{sec:Prompt Gamma-ray Emission and Implications for its Central Engine Activity}
\subsection{Two Prompt Emission Epochs with Distinct Temporal and Spectral Features}
We retrieve the \textit{Swift}/BAT data for GRB~240825A from the Swift Archive Download Portal\footnote{\url{https://www.swift.ac.uk/swift_portal/}} and extract light curves in the 15--25~keV, 25--50~keV, 50--100~keV, and 100--150~keV energy bands with a 128~ms time resolution. The upper-left panel of Figure~\ref{Fig2:Prompt} shows the light curve in the 15--50~keV band and the hardness ratio (HR), defined as the ratio of the fluxes in the 50--100 and 15--25 keV bands, $\mathrm{HR}=F_{50-100 {\rm\, keV}}/F_{15-25 {\rm\, keV}}$. The prompt emission evolves from an initial plateau to a rapid rise at $\sim 1$~s, followed by a steep decay after $\sim 1.55$~s. During the plateau phase, both the isotropic luminosity and HR remain stable ($L_{\gamma, \rm iso} \sim 1.50 \times 10^{50}$~erg s$^{-1}$ and $\mathrm{HR}\sim 0.98$). The subsequent rapid rise is accompanied by a steep luminosity increase ($L\propto t^{\sim 10}$) and significant spectral hardening, with HR increasing from 0.93 to 3.50. After the peak, the luminosity decays as a power law with intermittent flares, while the spectrum softens to $\mathrm{HR}\sim1.10$. At the late stage ($t>3$ s), the luminosity follows a power-law decay with $L\propto t^{-5/3}$, while the HR remains nearly constant. The second emission episode is also detected with the Fermi/GBM and Konus/Wind. The GBM light curve in the 50--300~keV band shows several overlapping short pulses lasting $\sim5$~s. The time-integrated GBM (8--1000 keV) spectrum from $T_0+0.96$ to $T_0+6.85$ s and Konus/Wind (20 keV--15 MeV) spectrum from $T_0$ to $T_0+28.16$ s are both well fitted with a Band function, yielding $E_{\rm p}=442\pm 8$ keV, $\alpha = -0.82\pm 0.01$, and $\beta=-2.22\pm 0.02$ for GBM \citep{2024GCN.37301....1S}, and $E_{\rm p}=403^{+40}_{-37}~\mathrm{keV}$, $\alpha=-0.79^{+0.05}_{-0.04}$, and $\beta=-2.10^{+0.06}_{-0.80}$ for Konus/Wind \citep{2024GCN.37302....1F}. The joint GBM and LAT spectrum in $T_0+0.53$--$T_0+5.83$~s further reveals a bimodal structure with an additional high-energy component peaking at $\sim0.1$~GeV \citep{2025ApJ...984L..45Z}. These results indicate that the early emission of the second episode is luminous and spectrally hard. 

\subsection{A Two-Step Central Engine Model for GRB 240825A}
\label{ssec: A Two-Step Central Engine model for GRB 240825A}
The distinct temporal and spectral properties of the two prompt-emission episodes motivate a possible two-stage central-engine scenario for GRB~240825A. In rotating core-collapse models, polar outflows can coexist with continued accretion through equatorial regions, allowing the PNS to gain mass and angular momentum \citep{2020MNRAS.492.4613O}. Material with sufficient angular momentum may circularize into a disk or torus, with the rate actually accreted by the PNS depending on angular-momentum transport and the interaction between the accretion flow and the magnetosphere. Continued mass growth, together with the evolution of thermal and rotational support, may eventually render the PNS unstable and lead to BH formation \citep{2021MNRAS.500.4365A}. A newly formed PNS typically has a mass of \(1.2\!-\!1.6\,M_\odot\), close to the effective Chandrasekhar mass of its progenitor's iron core \citep{1996ApJ...457..834T, 2002RvMP...74.1015W}. If the cold, nonrotating mass limit \(M_{\rm TOV}\approx2\!-\!2.6\,M_\odot\) is used as a reference, accretion of roughly one solar mass may bring the PNS closer to collapse. The actual collapse threshold also depends on the thermal and rotational support of the PNS \citep{2011ApJ...730...70O}. The collapse time can thus be estimated as $t_{\rm BH}\sim M_\odot/\dot M_{\rm acc,NS}$. Characteristic accretion rates of \(0.1\!-\!1\,M_\odot\,{\rm s}^{-1}\) allow substantial mass growth on timescales of seconds \citep{2021MNRAS.500.4365A, 2024ApJ...976L..13G}. This estimate motivates the proposed evolutionary channel but does not determine the pre-collapse accretion history or PNS lifetime of GRB~240825A. Before collapse, the proto-magnetar may power the early emission through a magnetized outflow. Because accretion can affect its spin evolution and electromagnetic spin-down power \citep{2018ApJ...857...95M}, we use the standard magnetic-dipole luminosity scaling only as an approximate description of the early plateau. The electromagnetic spin-down luminosity is estimated as $L_{\rm k}(t)=L_{\rm k,0}(1+t/\tau)^{-2}$ and $L_{\rm k,0}=10^{49}B_{0,15}^{2}P_{0,-3}^{-4}R_6^{6}\ {\rm erg\,s^{-1}}$, where $B_0$ is the initial surface dipole magnetic field and $\tau$ is the characteristic spin-down timescale (e.g. \citealt{2001ApJ...552L..35Z}). We attribute the early prompt emission plateau (the early epoch) of GRB 240825A to the dipole radiation at $t\ll\tau$ in a constant emitting efficiency in the BAT band ($\eta_{\rm md}\equiv L_{\rm BAT,1}/L_{\rm k,0}$). Thus, $B_0$ can be estimated with $B_0\simeq3.16\times10^{15}{\rm G} \ (L_{\rm BAT,1}/10^{49}\ {\rm erg\,s^{-1}})^{1/2}(\eta_{\rm md}/0.1)^{-1/2}(P_0/1\,{\rm ms})^2(R/10\,{\rm km})^{-3}$. Selecting $\eta_{\rm md}=0.01$, $0.1$, and $0.3$ as illustrative choices (e.g. \citealt{2021MNRAS.508.2505Z}), we have $B_0=(0.73\sim 3.97)\times10^{16}\ {\rm G}$ by taking $L_{\rm BAT,1}=1.58\times10^{50}\ {\rm erg\,s^{-1}}$, $P_0=1$~ms, and $R=10$~km.

Following the PNS-to-BH transition, the surrounding disk may help retain part of the magnetic flux generated during the PNS phase near the newly formed black hole. \citet{2024ApJ...976L..13G} proposed this inheritance channel and discussed the conditions under which the disk can confine the magnetic flux before it is lost through reconnection. This provides a possible origin for the magnetic flux required by the subsequent BZ-powered engine, although the flux transfer during collapse is not calculated in our model. The BH is subsequently fed by fallback material through an accretion disk. With fallback feeding and wind mass loss included, the conservation of mass and angular momentum in the accretion disk can be described as \citep{2008bhad.book.....K}:

\begin{equation}
\label{eq:diskmass}
\frac{\partial\Sigma}{\partial t}
=
\frac{1}{2\pi r}\frac{\partial\dot{M}}{\partial r}
+\dot{\Sigma}_{\mathrm{fb}}
-\dot{\Sigma}_{\mathrm{w}},
\end{equation}
\begin{equation}
\label{eq:ang}
\begin{aligned}
&\frac{\partial}{\partial t}\!\left(\Sigma\ell\right)
+\frac{1}{r}\frac{\partial}{\partial r}\!\left(r\Sigma v_{r}\ell\right)\\
&\qquad=
\frac{1}{r}\frac{\partial}{\partial r}\!\left(
r^{3}\nu\Sigma\frac{\partial\Omega}{\partial r}
\right)
+\ell_{\rm fb}\,\dot{\Sigma}_{\mathrm{fb}}
-\ell_{w}\,\dot{\Sigma}_{\mathrm{w}},
\end{aligned}
\end{equation}

Here $\Sigma$ is the disk surface density and $v_r$ is the radial velocity. The quantity $\dot{\Sigma}_{\rm fb}$ denotes the rate of increase in the disk surface density due to fallback material, while $\dot{\Sigma}_{\rm w}$ denotes the rate of decrease due to wind mass loss. The local specific angular momentum is $\ell=r^2\Omega$. We assume that the fallback material joins the disk with the local Keplerian specific angular momentum, $\ell_{\rm fb}(r,t)=\ell(r,t)$. The wind carries the local specific angular momentum at its
launch radius, $\ell_{\rm w}=f_j\ell$ with $f_j=1$, without an additional magnetic lever-arm torque \citep{2008MNRAS.388.1729K}. Assuming Keplerian rotation and neglecting changes in the orbital specific angular momentum caused by black-hole (BH) mass growth, we obtain the standard viscous relation for the accretion rate $\dot{M}$ (e.g. Equation~3.15 of \citealt{2008bhad.book.....K}):

\begin{equation} 
\label{eq:accretion}
\dot{M}(r,t)
=
6\pi\sqrt{r}\,\frac{\partial}{\partial r}\!\left(\nu\Sigma\,r^{1/2}\right),
\qquad
\nu=\alpha c_s H,
\end{equation}
where \(\nu\) is the effective kinematic viscosity, \(\alpha\) is the viscosity parameter, \(c_s\) is the sound speed of gas, and \(H\) is the half-thickness of the disk at radius $r$. The values of $c_{\rm s}$ and $H$ are determined by the advection factor $f_{\rm adv}(=q^{-}_{\rm adv}/q^{+}_{\rm adv})$ of the accretion flow, i.e. $c_{\rm s} \sim v_{\phi}\sqrt{f_{\rm adv}}$ and $h =H/r \approx \sqrt{f_{\rm adv}}$, where $q^{-}_{\rm adv}$ and $q^{+}_{\rm adv}$ are the factors of the advection cooling and viscous heating in the accretion flow, and $v_{\phi}$ is the Kepler's rotation velocity of gas around BH. To simplify the analysis, we adopt a fixed $h$ in our calculations \citep{2013PhRvD..87h4053S}.

Although the detailed wind-launching mechanism remains uncertain, disk winds can remove a substantial fraction of inflowing material, causing the net accretion rate to decrease toward smaller radii
\citep{1999MNRAS.303L...1B,2008MNRAS.388.1729K, 2018ApJ...854...43H}. Motivated by the radius-dependent inflow--outflow prescriptions
of \citet{2008MNRAS.388.1729K} and \citet{2018ApJ...854...43H}, we describe the wind strength using an effective index $s$, with $0\leq s\leq1$. In a source-free disk region approaching a quasi-steady state, the inward accretion rate satisfies

\begin{equation}
\frac{d\ln\dot{M}(r)}{d\ln r}=s.
\label{eq:wind_radial}
\end{equation}
In this limit, considering an annulus extending from $r$ to $r+dr$, the wind mass loss from the annulus equals the difference between the inward mass fluxes through its outer and inner boundaries, i.e. 
\(
d\dot{M}_{\rm w}
=
\dot{M}(r+dr)
-
\dot{M}(r)
\simeq
\frac{d\dot{M}}{dr}\,dr 
\). Thus, we have

\begin{equation}
d\dot{M}_{\rm w} =
\frac{s}{r}\dot{M}(r)\,dr.
\label{eq:annulus_wind}
\end{equation}
For the time-dependent disk calculation, we use this annular relation as a local closure. The total wind mass-loss rate from an annulus of width $dr$ is \(d\dot{M}_{\rm w}=
2\pi r\,dr\,\dot{\Sigma}_{\rm w}\), where $\dot{\Sigma}_{\rm w}$ is the surface-density sink, which is

\begin{equation}
\dot{\Sigma}_{\rm w}(r,t)
=
\frac{s}{2\pi r^2}
\dot{M}_{\rm inflow}(r,t),
\label{eq:wind_sink}
\end{equation}
in which 
$\dot{M}_{\rm inflow}(r,t)\equiv \max[\dot{M}(r,t),0]$ is the instantaneous local inward viscous mass flux. Equation~(\ref{eq:wind_sink}) is an instantaneous local extension of the quasi-steady inflow and outflow prescription. It is not an additional constraint on $\partial\dot{M}/\partial r$ in the full time-dependent disk. Fallback injection and time-dependent disk storage remain explicit terms in the mass-conservation equation. Equation~(\ref{eq:wind_radial}) is therefore recovered only in the source-free, quasi-steady limit.

The material is assumed to fall onto the disk near $r_{\rm fb}$, hereafter referred to as the fallback radius. We model the fallback source as a Gaussian profile, i.e. \citep{2016MNRAS.463..245L, 2025ApJ...985..195L}

\begin{equation}
\label{eq:Sigmafb}
\dot{\Sigma}_{\rm fb}(r,t)
=
\frac{\dot{M}_{\rm fb}(t)}{2\pi r r_{\rm fb}}
A
\exp\left[
-\left(
\frac{r-r_{\rm fb}}
{r_{\rm fb}/4}
\right)^2
\right]
\end{equation}
The dimensionless constant $A$ normalizes the fallback source term such that $\int_{r_{\rm in}}^{r_{\rm out}} 2\pi r\,\dot{\Sigma}_{\rm fb}(r,t)\,dr=\dot{M}_{\rm fb}(t)$. The fallback rate $\dot{M}_{\rm fb}$ describes the mass supplied to the disk per unit time and is parameterized as \citep{2018ApJ...857...95M}

\begin{equation}
\dot{M}_{\mathrm{fb}}=\frac{2}{3}\frac{M_{\rm fb}}{t_{\rm fb}}
\left(1+t/t_{\rm fb}\right)^{-5/3},
\end{equation}
where $M_{\rm fb}$ is the total mass supplied to the disk during the modeled post-collapse phase and $t_{\rm fb}$ is the characteristic mass-supply timescale, hereafter referred to as the characteristic fallback timescale. Although the black hole lacks an intrinsic dipolar magnetic field, large-scale magnetic flux advected by the fallback-fed disk can thread the event horizon and extract the rotational energy from the BH spin through the Blandford--Znajek mechanism \citep{1977MNRAS.179..433B, 2000PhR...325...83L}. We estimate the strength of this field by assuming approximate balance between the magnetic pressure near the horizon and the ram pressure of the innermost accretion flow \citep{2011MNRAS.418L..79T, 2017ApJ...849...47L}:

\begin{equation}
\frac{B_{\text{BH}}^2}{8\pi} = P_{\text{ram}} \sim \rho c^2 \sim \frac{\dot{M}(r_{\text{in}}) c}{4\pi r_{\text{H}}^2},
\end{equation}
Here, $r_{\rm H}=(1+\sqrt{1-a^2})r_g$ is the BH horizon radius, and $r_g=GM_{\rm BH}/c^2$ is the gravitational radius. The constants $G$ and $c$ denote the gravitational constant and the speed of light, respectively. The inner accretion rate $\dot{M}(r_{\rm in})$ is calculated from Eq.~\eqref{eq:accretion}. The resulting BZ power is expressed as\citep{2022ApJ...936L..10Z}

\begin{equation}
\label{eq:LBZ}
     L_{\mathrm{BZ}} = 9.3 \times 10^{53} \, \mathrm{erg \, s^{-1}} \,
     \frac{a^{2} F(a)}
     {\left(1 + \sqrt{1 - a^{2}}\right)^{2}} \,
     \frac{\dot{M}(r_{\rm in})}{M_{\odot} \, \mathrm{s^{-1}}},
\end{equation}
where $a$ is the dimensionless BH spin parameter and $r_{\rm in}$ is the disk inner radius, which we identify with the innermost stable circular orbit (ISCO). The function $F(a)$ is defined as $F(a)=[(1+q^{2})/q^{2}][(q+1/q)\arctan q-1]$, where $q=a/(1+\sqrt{1-a^{2}})$.

The BH mass and angular momentum evolve as \citep{2000PhR...325...83L,2000JKPS...36..188L}

\begin{align}
\frac{dM_{\mathrm{BH}}}{dt} &= \dot{M}(r_{\mathrm{in}}) e_{\mathrm{ms}} - \frac{L_{\mathrm{BZ}}}{c^2}, \label{eq:bhmass} \\
\frac{dJ_{\mathrm{BH}}}{dt} &= \dot{M}(r_{\mathrm{in}}) l_{\mathrm{ms}} - \frac{2L_{\mathrm{BZ}}}{\Omega_{\mathrm{H}}}, \label{eq:bhspin}
\end{align}
where $J_{\mathrm{BH}}=GM_{\mathrm{BH}}^2a/c$ is the BH angular momentum and $\Omega_{\mathrm{H}}=ca/(2r_{\mathrm{H}})$ is the angular velocity of the horizon. The quantities $e_{\mathrm{ms}}$ and $l_{\mathrm{ms}}$ denote the specific energy in units of $c^2$ and the specific angular momentum at the marginally stable orbit, respectively, and are given by $e_{\mathrm{ms}}=(3\hat{r}_{\mathrm{ms}})^{-1/2}(4-3a/\sqrt{\hat{r}_{\mathrm{ms}}})$ and $l_{\mathrm{ms}}=(2\sqrt{3}GM_{\mathrm{BH}}/c)(1-2a/(3\sqrt{\hat{r}_{\mathrm{ms}}}))$. Here, $\hat{r}_{\mathrm{ms}}=r_{\mathrm{ms}}/r_g$ is the dimensionless orbital radius, given by $\hat{r}_{\mathrm{ms}}=3+Z_2-[(3-Z_1)(3+Z_1+2Z_2)]^{1/2}$, with $Z_1=1+(1-a^2)^{1/3}[(1+a)^{1/3}+(1-a)^{1/3}]$ and $Z_2=(3a^2+Z_1^2)^{1/2}$ \citep{1972ApJ...178..347B,2008bhad.book.....K}. 

\begin{figure}[htbp]
     \centering
     \includegraphics[width=0.49\textwidth, angle=0]{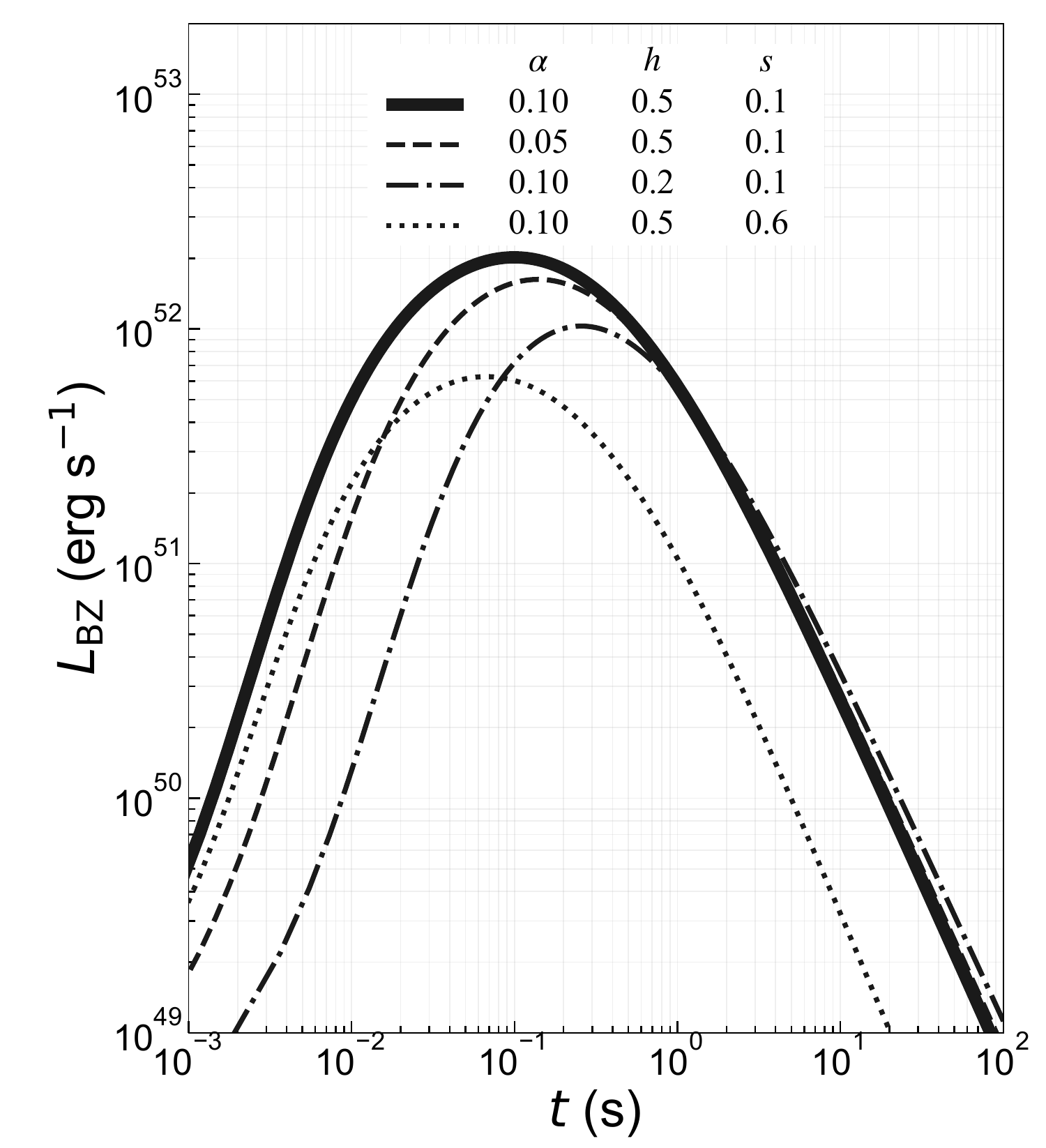}
     \includegraphics[width=0.49\textwidth, angle=0]{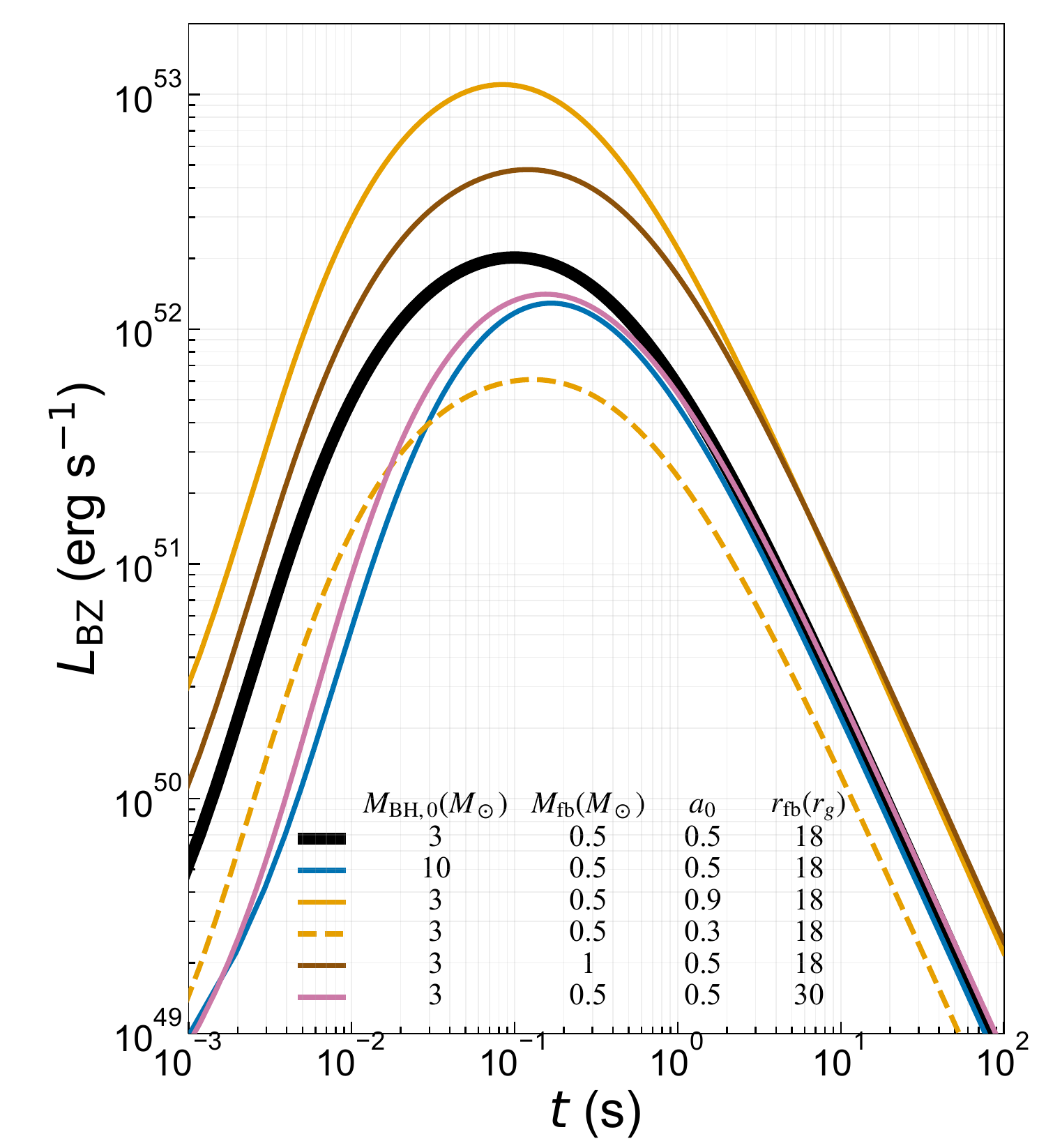}
     \caption{Illustrations of the parameter dependence on the BZ luminosity evolution. {\em Left panel---} Dependence of $L_{\rm BZ}$ on the disk parameters $(\alpha,h,s)$, with the initial black-hole conditions and fallback-supply parameters fixed at $(M_{\rm BH,0},a_0,M_{\rm fb},r_{\rm fb})=(3\,M_\odot,0.5,0.5\,M_\odot,18r_g)$. {\em Right panel---} Dependence of $L_{\rm BZ}$ on the initial black-hole conditions and fallback-supply parameters $(M_{\rm BH,0},a_0,M_{\rm fb},r_{\rm fb})$, with the disk parameters fixed at $(\alpha,h,s)=(0.10,0.5,0.1)$. The characteristic fallback timescale is fixed at $t_{\rm fb}=0.2\,{\rm s}$ in both panels.}
\label{Fig1:parameter_dependence}
\end{figure}

Figure~\ref{Fig1:parameter_dependence} illustrates the temporal evolution of $L_{\rm BZ}$ adopting various sets of the accretion disk and black hole parameters. The characteristic mass-supply timescale is fixed at $t_{\rm fb}=0.2$ s, while the peak time of $L_{\rm BZ}$ also depends on the disk response to this supply. The peak luminosity is $10^{51}\sim 10^{53}$ erg/s, which is comparable to the geometrically-corrected GRB jet luminosity ($\sim 10^{51}$ erg/s; \citealt{2001ApJ...562L..55F, 2003ApJ...590..379B}) assuming a radiative efficiency of $\eta_\gamma=0.1\sim 0.01$. As shown in the left panel of Figure~\ref{Fig1:parameter_dependence}, decrease of the viscosity parameter $\alpha$ or the disk half-thickness ratios $h$ leads to decrease of the peak luminosity and to delay the peak time, and increase of the wind parameter $s$ results in decrease of the peak luminosity, with the initial black-hole conditions and fallback-supply parameters fixed at $(M_{\rm BH,0},a_0,M_{\rm fb},r_{\rm fb},t_{\rm fb})=(3\,M_\odot,0.5,0.5\,M_\odot,18r_g,0.2\,{\rm s})$. Fixing the disk parameters at $(\alpha,h,s)=(0.10,0.5,0.1)$ and setting $t_{\rm fb}=0.2\,{\rm s}$, we find that the peak BZ luminosity is sensitive to the initial BH spin ($a_0$) and the fallback mass ($M_{\rm fb}$), and the peak time depends on the fallback radius $r_{\rm fb}$, as shown in the right panel of Figure~\ref{Fig1:parameter_dependence}.

\subsection{Representative Initial BH Conditions and Fallback-Accretion Modeling}
Interpreting the second episode as emission from a BZ-powered jet, we relate the observed gamma-ray luminosity to the BZ jet power as follows:

\begin{equation}
f_{\rm b} k L_{\gamma,\rm iso}
=
\eta_{\gamma} L_{\rm BZ},
\label{eq:LobsLBZ}
\end{equation}
where $\eta_{\gamma}$ denotes the efficiency for converting the BZ jet power into prompt gamma-ray emission in the $1-10^4$ keV band, $k$ is the $k$-correction factor to correct the observed fluence in the BAT band to the $1-10^4$ keV band in the source frame, and $f_b$ is the beaming correction factor. We set $\eta_\gamma$ and $k$ as constant during the burst. 
We adopt $\eta_\gamma=0.01$ as a fiducial conversion efficiency. A value of this order has been used in fallback-powered X-ray modeling \citep{2013ApJ...767L..36W}, although its applicability to prompt gamma-ray emission remains uncertain. Adopting the spectral parameters derived from the fit to the GBM data, we have $k=6.5$. Our afterglow modeling gives a characteristic jet half-opening angle of $\theta_j\simeq6.4^\circ$ (see Table \ref{tab:table3}), which corresponds to $f_{\rm b}\simeq6.3\times10^{-3}$. The radiative efficiency and beaming factor determine the conversion between the modeled BZ power and the observed luminosity, with the main uncertainty originating in the unknown gamma-ray radiative efficiency.

\begin{figure}[htbp]
     \centering
     \includegraphics[width=0.49\textwidth, angle=0]{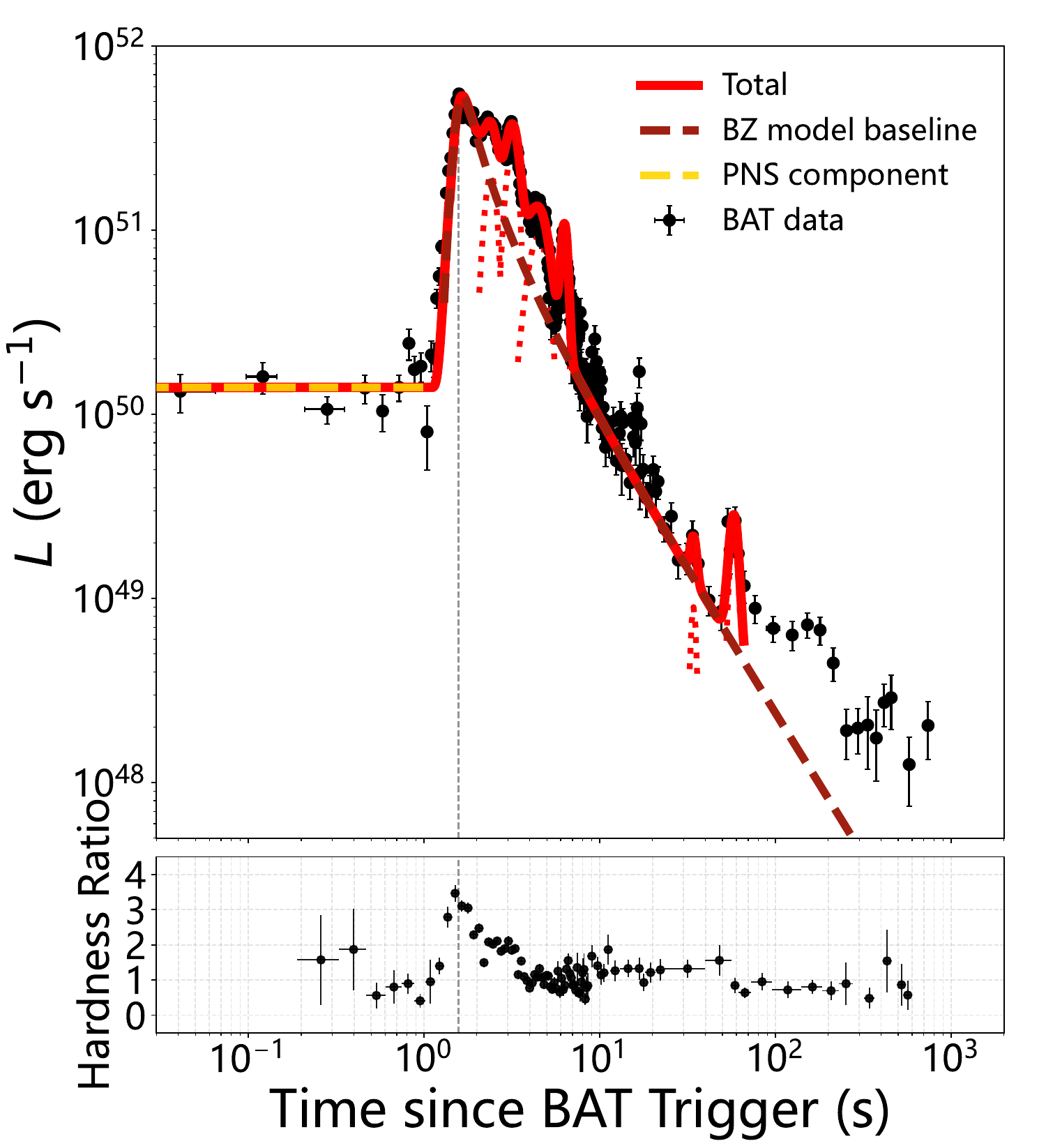}
     \includegraphics[width=0.49\textwidth, angle=0]{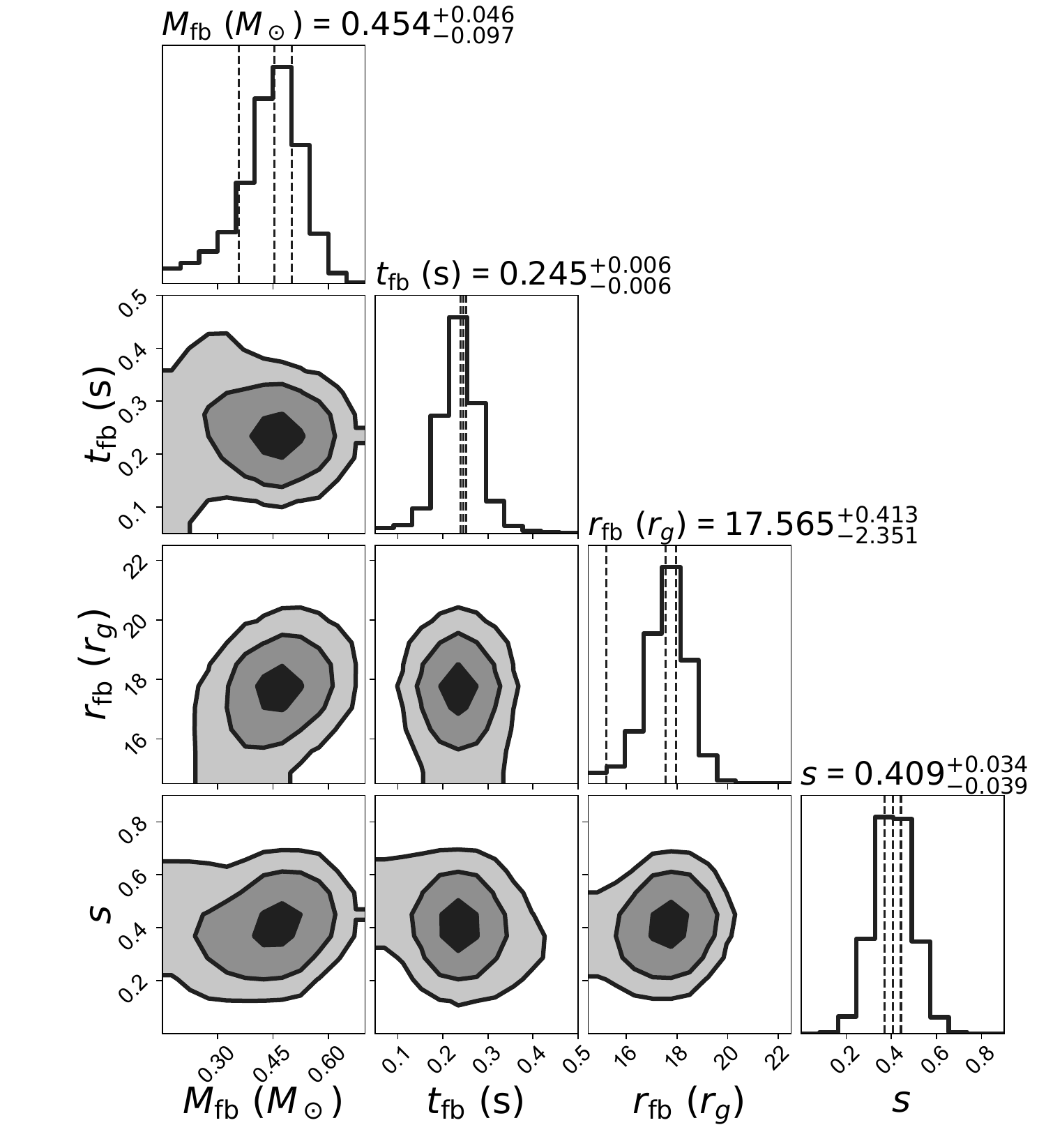} \\
     
     \includegraphics[width=0.23\textwidth, angle=0]{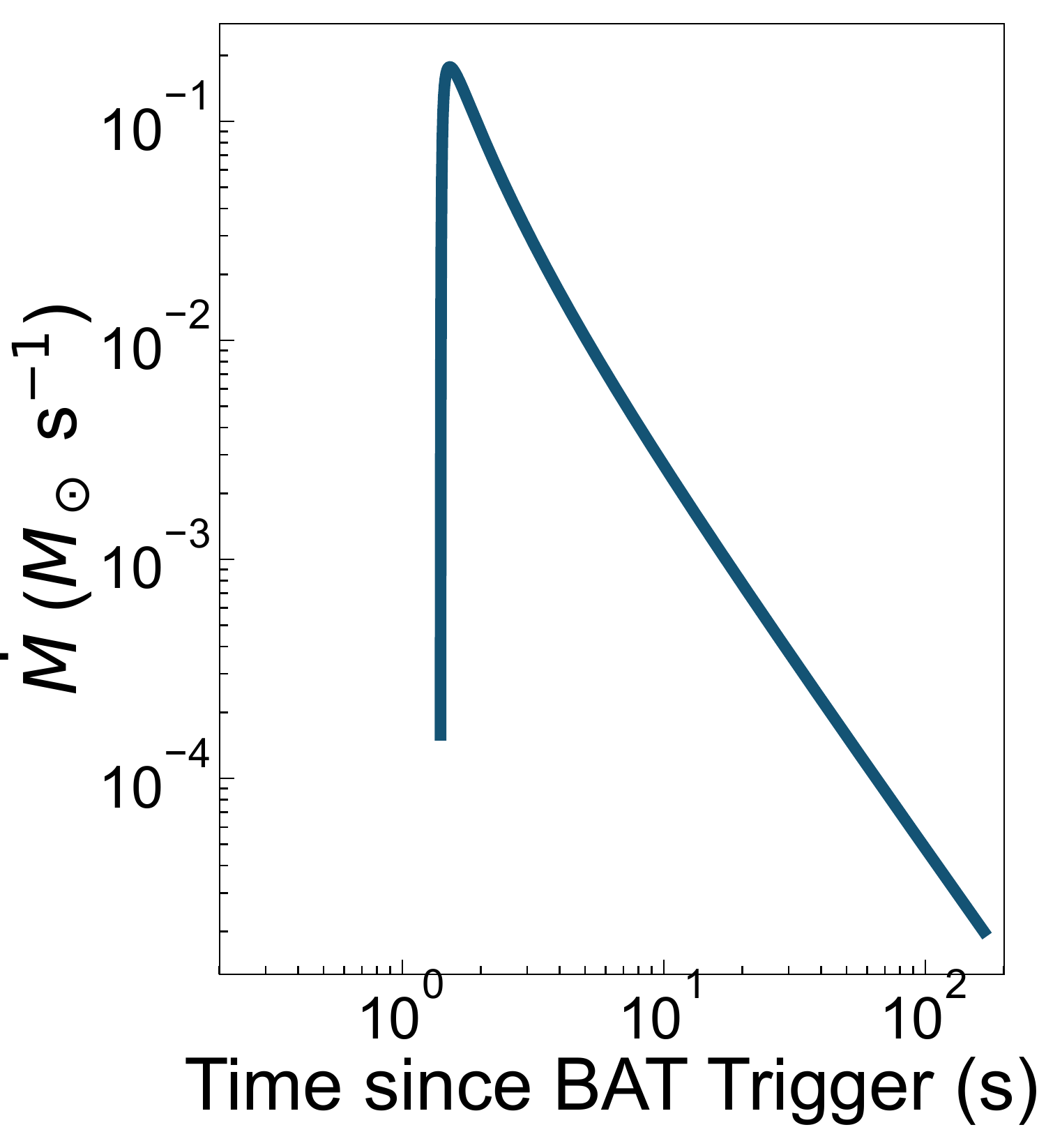}
     \includegraphics[width=0.23\textwidth, angle=0]{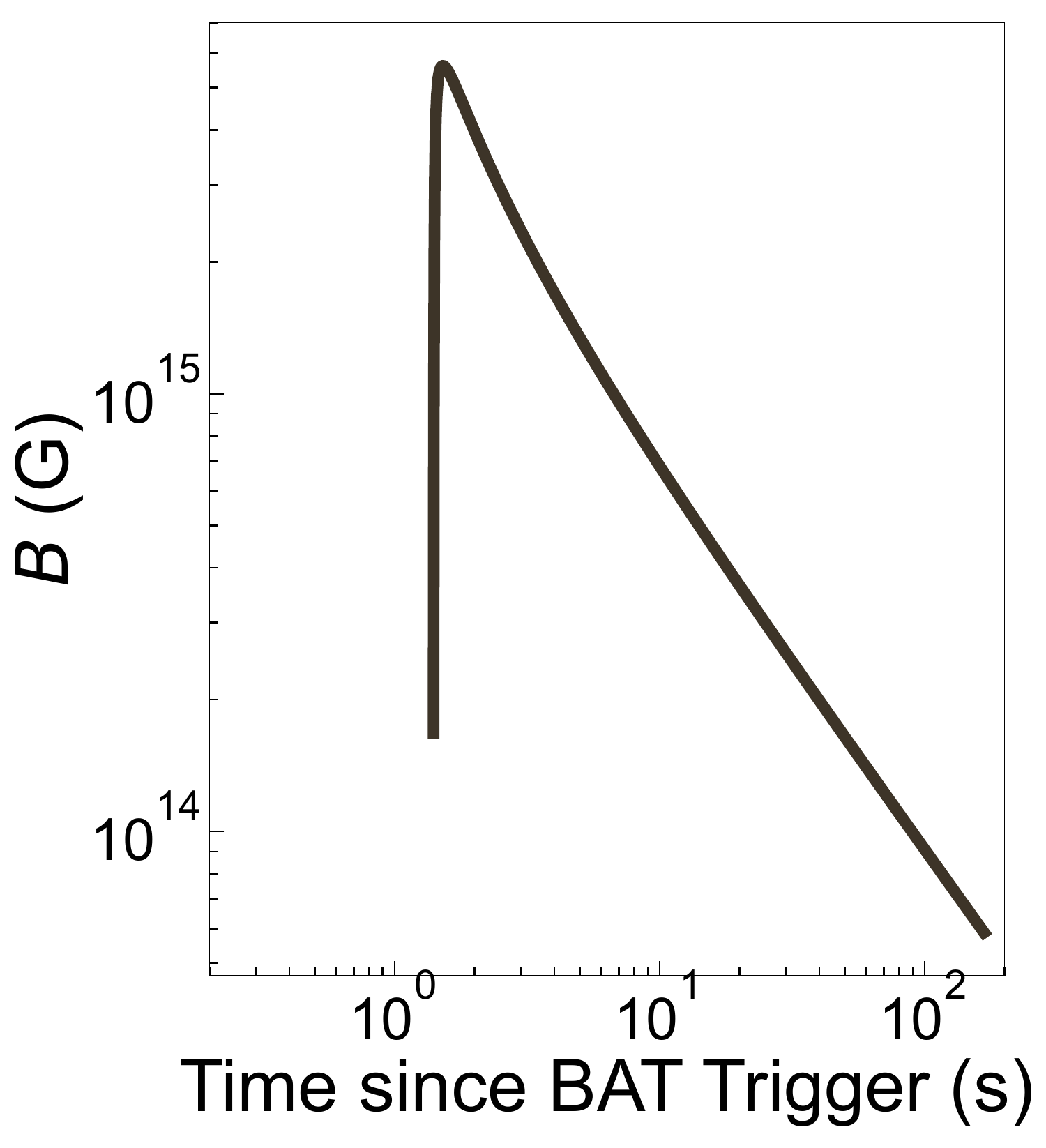}
     \includegraphics[width=0.23\textwidth, angle=0]{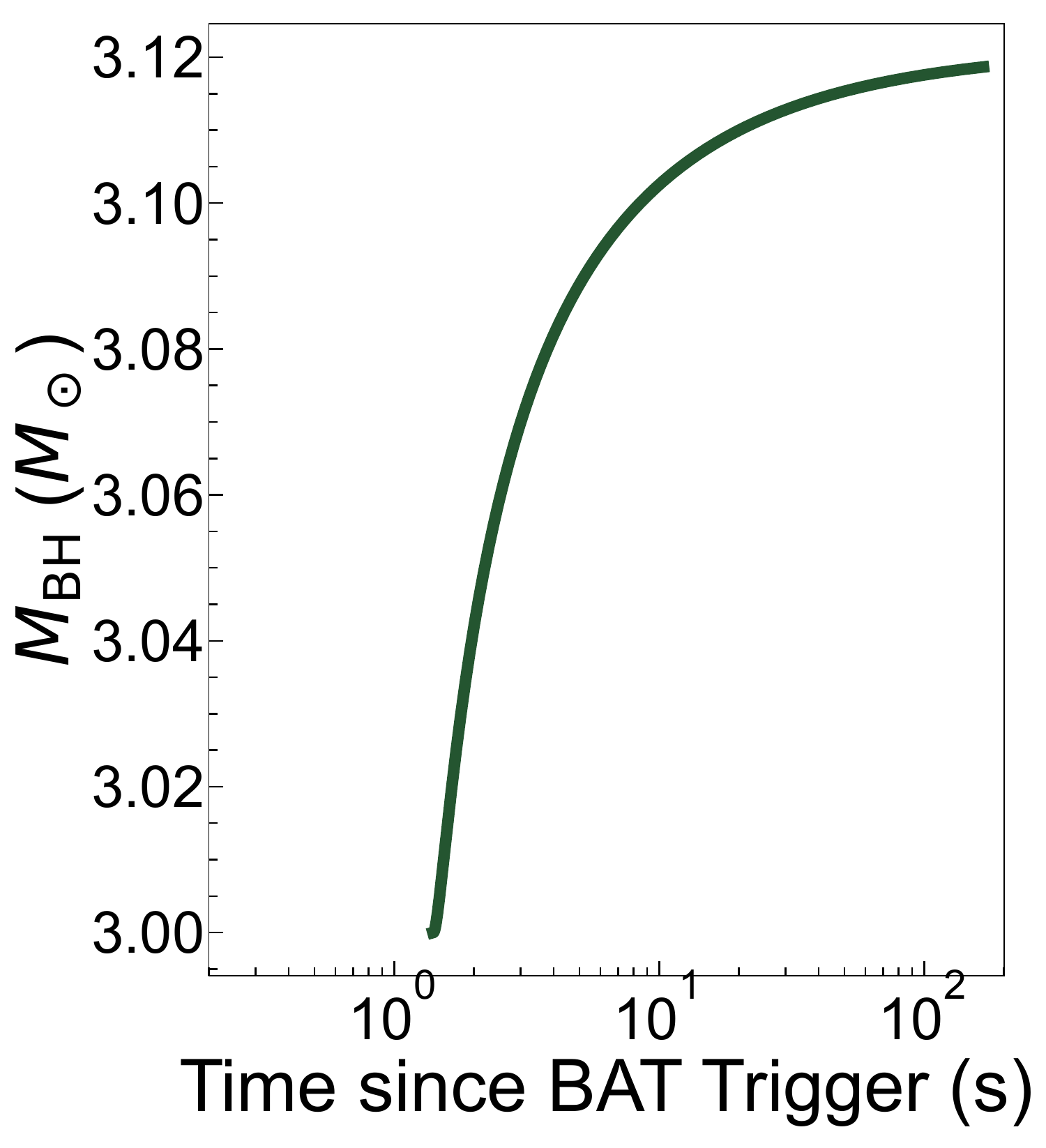}
     \includegraphics[width=0.23\textwidth, angle=0]{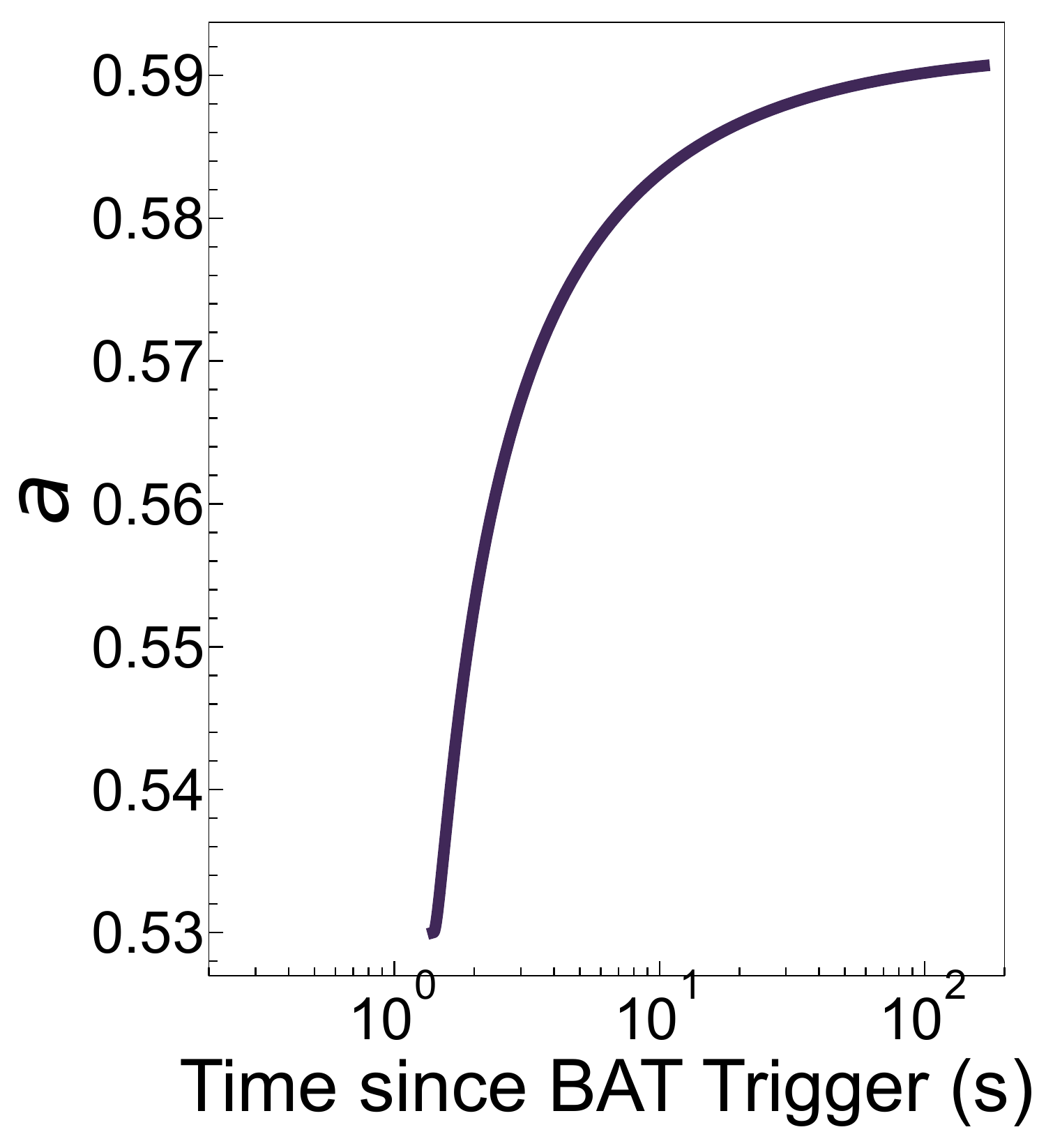}
     \caption{Modeling of GRB~240825A within the proposed two-stage central-engine scenario, with the initial BH mass and spin fixed at $(M_{\rm BH,0},a_0)=(3.0\,M_\odot,0.53)$. Upper-left: The \textit{Swift}/BAT 15--50~keV luminosity light curve and the corresponding hardness ratio. The solid red curve shows the total model, including the Gaussian pulse components. The yellow and brown dashed curves represent the proto-magnetar component and the BZ-powered baseline, respectively, while the dotted red curves show the individual Gaussian components. Upper-right: Posterior distributions of the fallback-accretion parameters $(M_{\rm fb},t_{\rm fb},r_{\rm fb},s)$ obtained from fitting the BAT light curve with the adopted initial BH configuration. Diagonal panels show the one-dimensional marginalized posterior distributions, with dashed lines marking the 16th, 50th, and 84th percentiles. Off-diagonal panels show the two-dimensional joint marginalized posterior distributions. Lower panels, from left to right: Evolution of the inner accretion rate $\dot{M}_{\rm in}$, the inferred horizon magnetic field strength $B$, the BH mass $M_{\rm BH}$, and the dimensionless spin $a$ for the corresponding fallback-accretion solution.}
\label{Fig2:Prompt}
\end{figure}

With these assumptions, we first model the later episode using a reference configuration with $(M_{\rm BH,0},a_0)=(3.0\,M_\odot,0.53)$. This choice is based on the estimate of \citet{2006ApJ...637..914W}, who predicted $a_0=0.53$ for a black hole formed from the inner $3\,M_\odot$ of their pre-supernova model 35OC. We set the transition time to $t_{\rm tr}\sim1.5\,{\rm s}$ in the observer frame, based on the observed hardness-ratio transition. The fallback-accretion phase starts at the corresponding source-frame time $t_{\rm tr}/(1+z)$, and the time in the fallback-rate prescription is measured from this onset. The upper-left panel of Figure~\ref{Fig2:Prompt} shows the resulting light-curve fit for the proposed two-stage central-engine scenario. With the initial BH mass and spin fixed, we obtain posterior estimates of $M_{\rm fb}=0.454^{+0.046}_{-0.097}\,M_\odot$, $r_{\rm fb}=17.565^{+0.413}_{-2.351}\,r_g$, $t_{\rm fb}=0.245^{+0.006}_{-0.006}\,{\rm s}$, and $s=0.409^{+0.034}_{-0.039}$. Flares during the decay phase are described phenomenologically by Gaussian components superposed on the BZ model baseline. The lower panels of Figure~\ref{Fig2:Prompt} show the evolution of the inner accretion rate $\dot{M}_{\rm in}$, the inferred horizon magnetic field $B$, the BH mass $M_{\rm BH}$, and the dimensionless spin $a$. As fallback proceeds, viscous transport regulated by advective cooling, disk-wind mass loss and viscous heating reshapes the initial narrow Gaussian surface-density profile into an inner-enhanced and outer-depleted configuration. Under the adopted pressure-balance assumption, the magnetic field strength near the horizon gradually decreases as the inner accretion rate $\dot{M}(r_{\rm in})$ declines.

The parameter-dependence analysis above shows that the modeled BZ luminosity varies with the initial BH conditions, especially the spin over the range considered. We therefore repeat the fits with additional initial BH masses and spins drawn from the core-collapse simulations of \citet{2021MNRAS.500.4365A}, using the PNS masses and dimensionless angular momenta just before collapse as estimates of the newborn BH values (their Table~2). Table~\ref{tab:table2} summarizes the adopted initial BH configurations and the posterior estimates of the fallback parameters, with the peak inner accretion rate $\dot{M}_{\rm in}^{\rm peak}$ derived from each fitted solution listed in the last column. Lower initial spins generally require higher peak accretion rates to reproduce the observed luminosity. The low-spin configuration from model 35OC-Sw serves only as an illustrative comparison, for which our disk model yields a formal fit with $\dot{M}_{\rm in}^{\rm peak}\simeq1.38\,M_\odot\,{\rm s^{-1}}$. Although some collapsar simulations yield accretion rates of this order \citep{2026PhRvD.113h3020I}, the original 35OC-Sw progenitor lacks sufficient angular momentum for early disk formation and is therefore unlikely to produce a standard collapsar engine \citep{2021MNRAS.500.4365A}. Excluding this illustrative low-spin case, the fitted central values span $M_{\rm fb}\simeq0.45$--$0.96\,M_\odot$, $r_{\rm fb}\simeq16.9$--$17.6\,r_g$, $t_{\rm fb}\simeq0.13$--$0.25\,{\rm s}$, and $s\simeq0.35$--$0.41$.

\section{Properties of Afterglow Jet and its Environment}
\label{sec:Properties of Afterglow Jet and its Environment}
\subsection{Time-dependent Properties of the burst environment}
\label{ssec:Time-dependent Properties of the burst environment}
The afterglow of GRB 240825A was followed up over an extended temporal baseline from radio to GeV gamma-ray bands. We compile the multi-wavelength afterglow observations \citep{2024GCN.37292....1S,2024GCN.37338....1S,2024GCN.37388....1G,2024GCN.37278....1Z,2024GCN.37293....1M,2024GCN.37288....1D,2025RAA....25j5003W}, and present the corresponding light curves in Figure~\ref{Fig3:Afterglow}. Galactic extinction correction is applied to the optical data using the dust maps of \citet{1998ApJ...500..525S}, recalibrated by \citet{2011ApJ...737..103S}. The corresponding Galactic color excess along the line of sight is $E_{B-V}^{\rm G}=0.053$, yielding $A^{\rm G}_{\rm V}=0.166$, $A^{\rm G}_{\rm R}=0.131$, and $A^{\rm G}_{\rm I}=0.094$~mag\footnote{\href{https://irsa.ipac.caltech.edu/applications/DUST/}{https://irsa.ipac.caltech.edu/applications/DUST/}}. The Galactic hydrogen column density is set to $N_{\rm H}^{\rm G} = 5.80 \times 10^{20}~\mathrm{cm^{-2}}$. We fit the optical-to-X-ray spectral energy distributions (SEDs) with the \texttt{Xspec} package \citep{1996ASPC..101...17A}, adopting a Milky Way-like extinction law for the host galaxy with $R_V = 3.1$ \citep{1992ApJ...395..130P}. The joint SEDs over three selected intervals, $T_0+[7.0\times10^{1},2.0\times10^{2}]~\mathrm{s}$, $T_0+[4.03\times10^{2},1.0\times10^{3}]~\mathrm{s}$, and $T_0+[4.0\times10^{3},2.0\times10^{4}]~\mathrm{s}$, are well fitted by an absorbed single power-law model. The best-fit results are presented in the right panel of Figure~\ref{Fig3:Afterglow}, with the corresponding parameters summarized in Table~\ref{tab:table1}. The photon indices indicate modest spectral softening over the three intervals, with $\Gamma = 1.53 \pm 0.01$, $1.57 \pm 0.01$, and $1.70 \pm 0.06$, respectively. In contrast, the inferred host extinction shows significant temporal variation, yielding $E_{B-V}^{\rm host} = 0.41 \pm 0.01$, $0.63 \pm 0.02$, and $0.37 \pm 0.08$ in the three intervals, accompanied by $N_{\rm H}^{\rm host}$ values of $(0.35 \pm 0.02) \times 10^{22}$, $(0.64 \pm 0.02) \times 10^{22}$, and $(0.56 \pm 0.07) \times 10^{22}~\mathrm{cm^{-2}}$, respectively. These results suggest temporal variations in the extinction and absorption properties along the line of sight. The temporal evolution of $E^{\rm host}_{\rm B-V}$ can be described by an empirical function, $E^{\rm host}_{B-V}=-0.304\,x^2+1.734\,x-1.842$, where $x=\log_{10}(t/{\rm s})$. The corresponding evolution is shown in the lower subpanel of Figure~\ref{Fig3:Afterglow} and follows a similar trend to the temporal behavior of $N^{\rm host}_{\rm H}$ reported by \citet{2026ApJ...997..101L}. 

\subsection{Properties of the Afterglow Jet}
\begin{figure}[htbp]
     \centering
     \includegraphics[width=0.45\textwidth, angle=0]{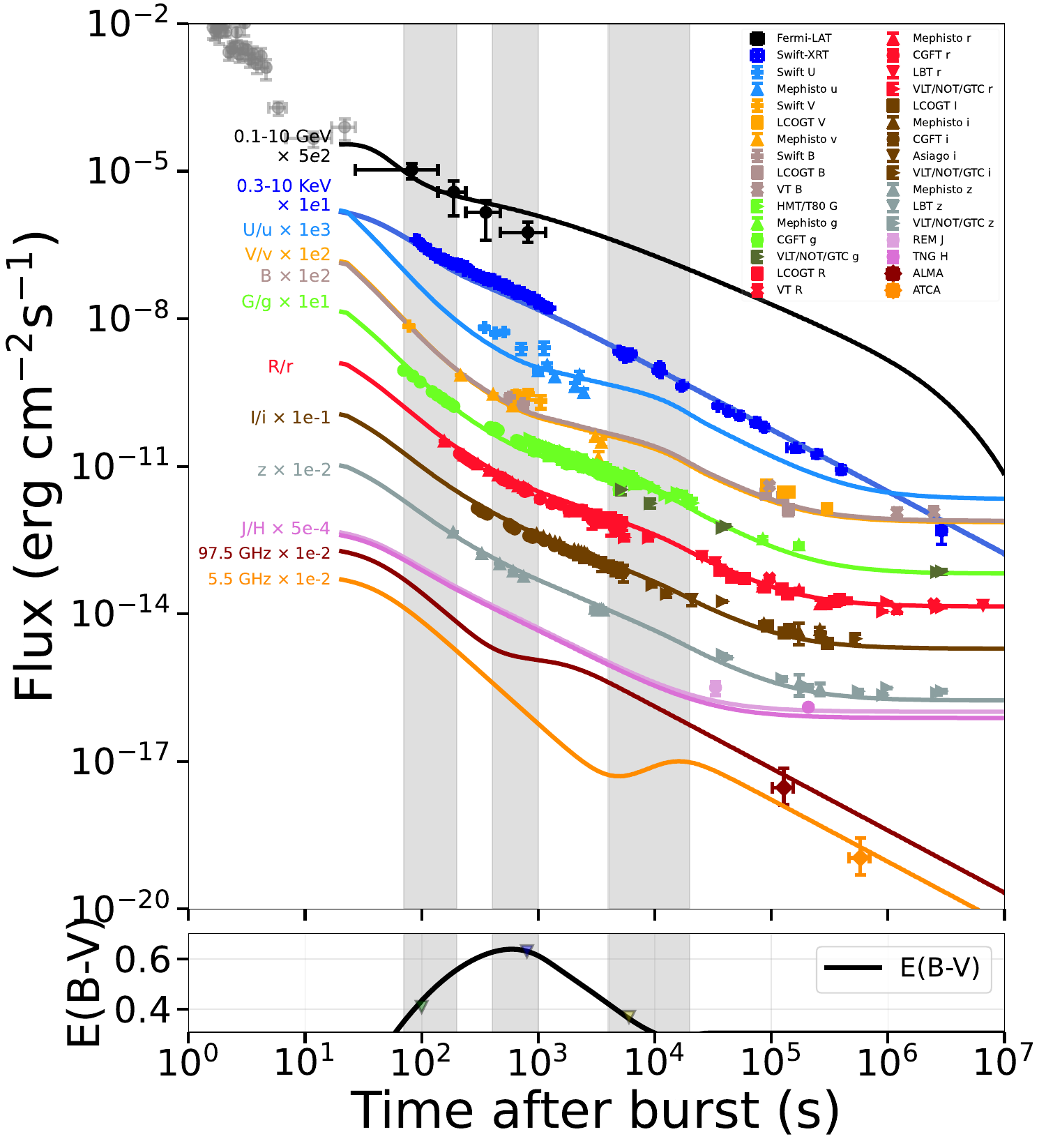}
     \includegraphics[width=0.45\textwidth, angle=0]{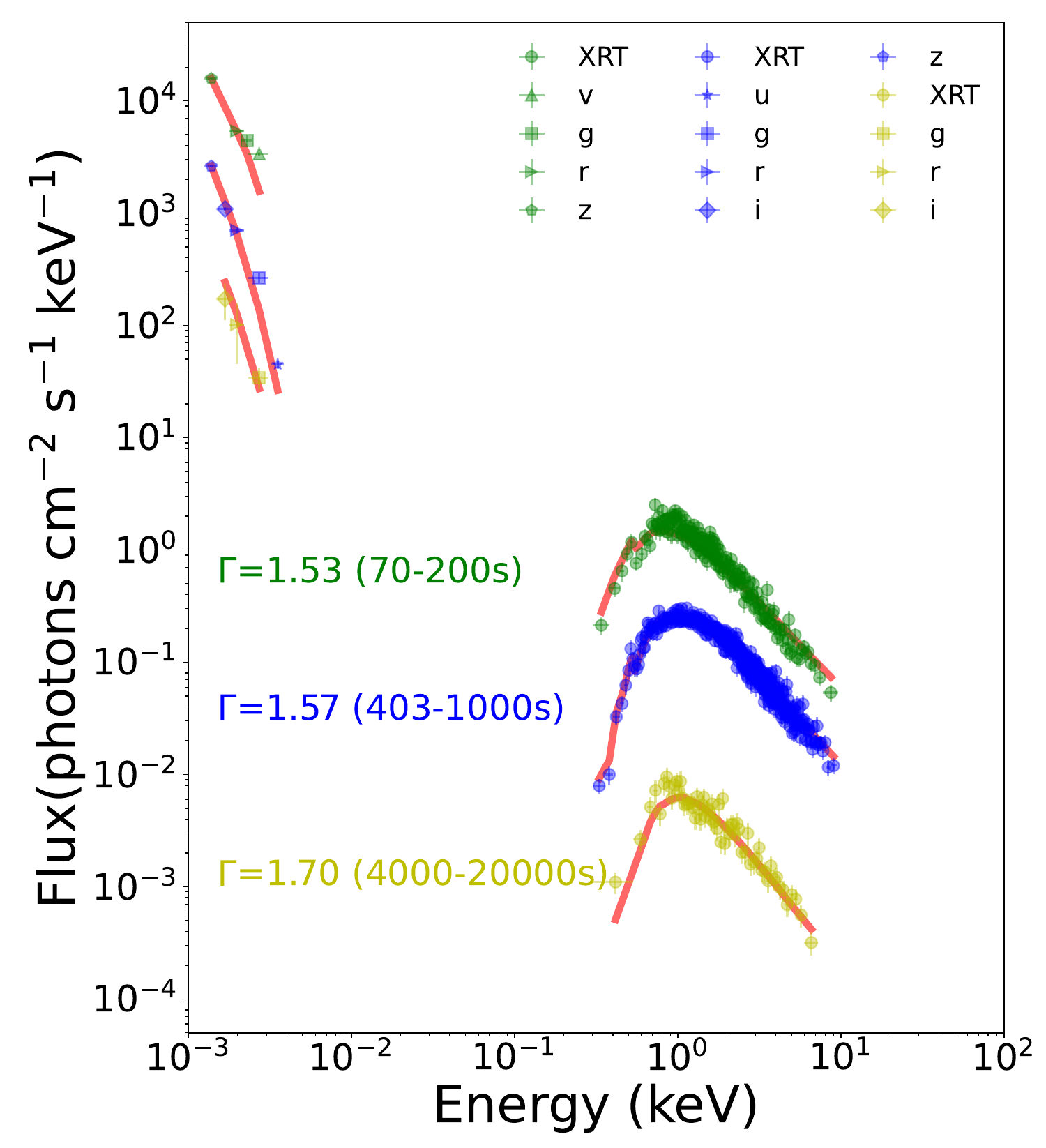}

     \includegraphics[width=0.29\textwidth, angle=0]{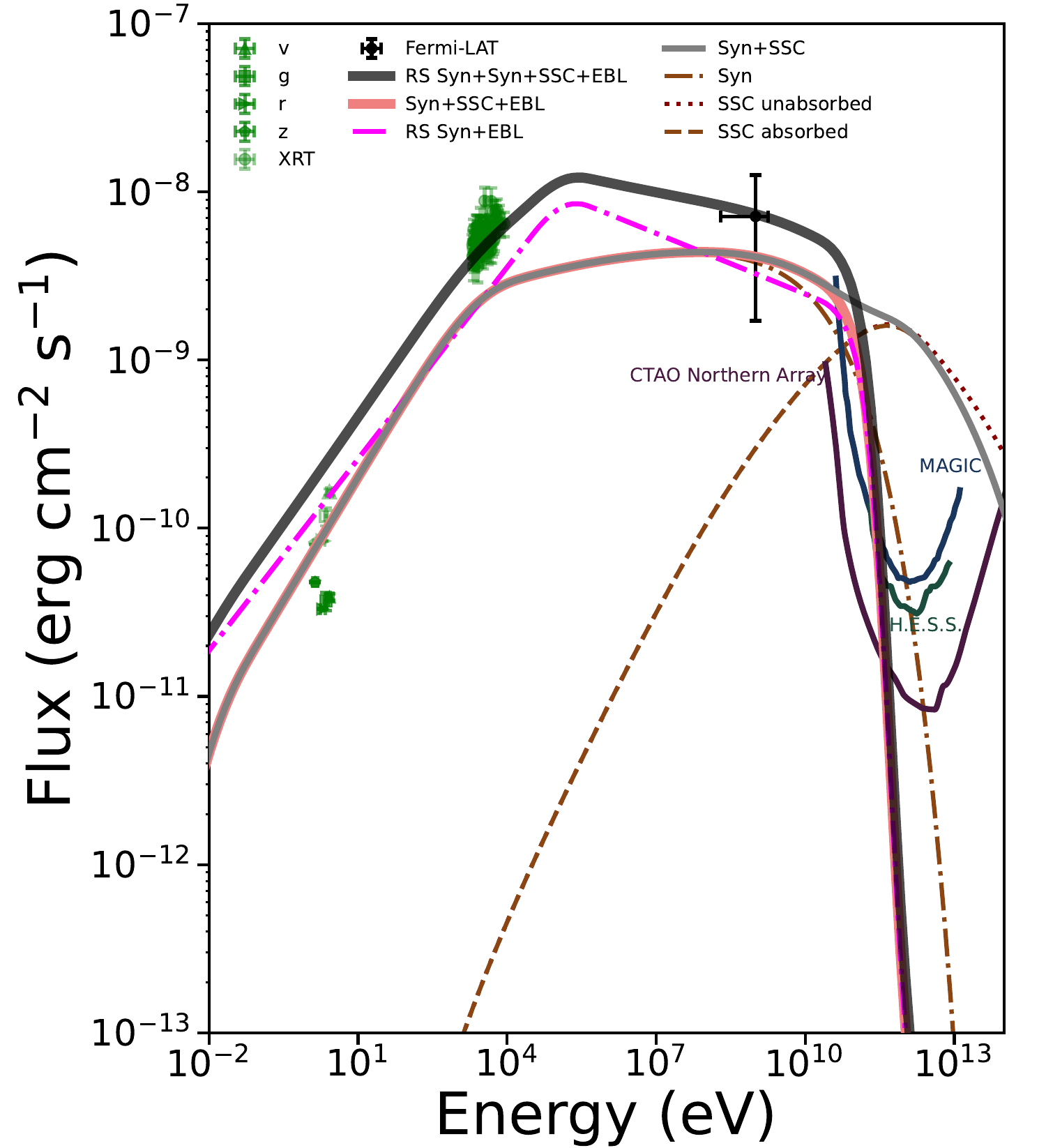}
     \includegraphics[width=0.29\textwidth, angle=0]{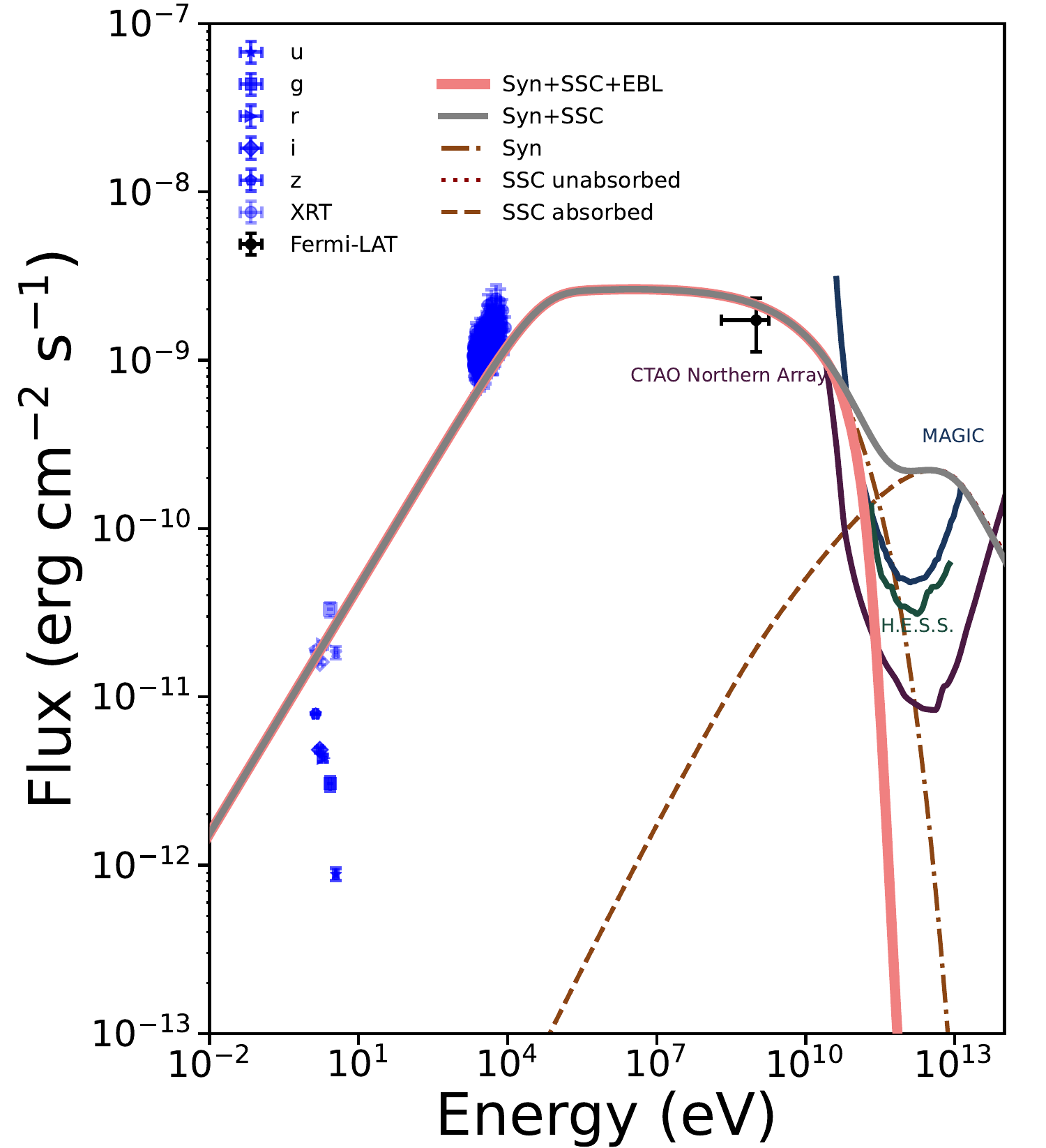}
     \includegraphics[width=0.29\textwidth, angle=0]{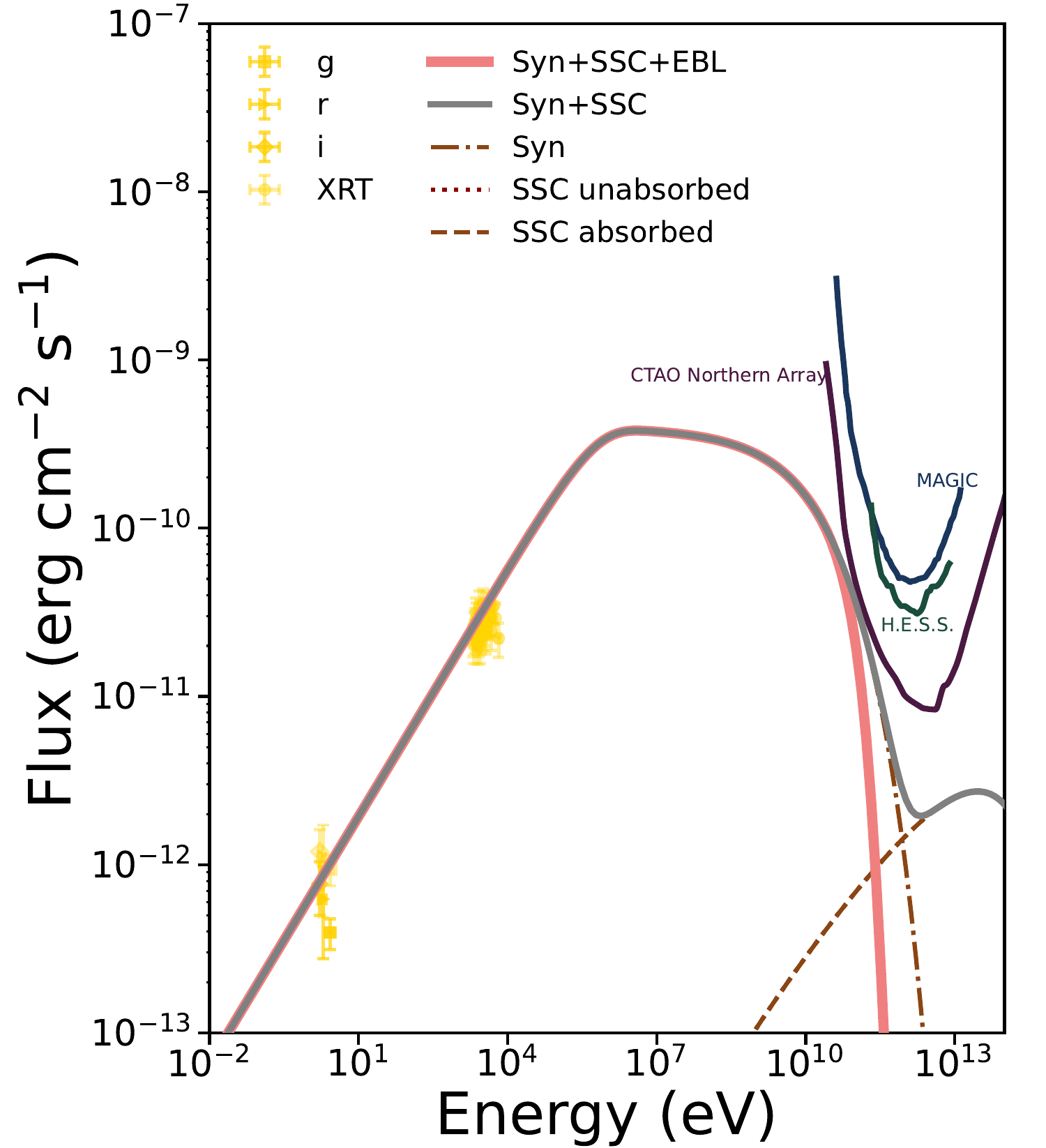}

     \caption{Upper-left panel: Multi-wavelength afterglow light curves of GRB~240825A and the evolution of the host extinction $E^{\rm host}_{B-V}$. The three colored point sets indicate the $E^{\rm host}_{B-V}$ values derived from joint spectral fitting. 
     Upper-right panel: Joint X-ray and optical spectral fits in three time intervals; solid curves show absorbed power-law models.
     Bottom panels: Time-integrated spectras for the three intervals. Filled/open symbols represent observed/extinction-corrected flux densities, respectively, with high-energy sensitivity limits from CTAO, MAGIC, and H.E.S.S.}
     \label{Fig3:Afterglow}
\end{figure}

\begin{figure}[htbp]
     \centering
     \includegraphics[width=0.6\textwidth, angle=0]{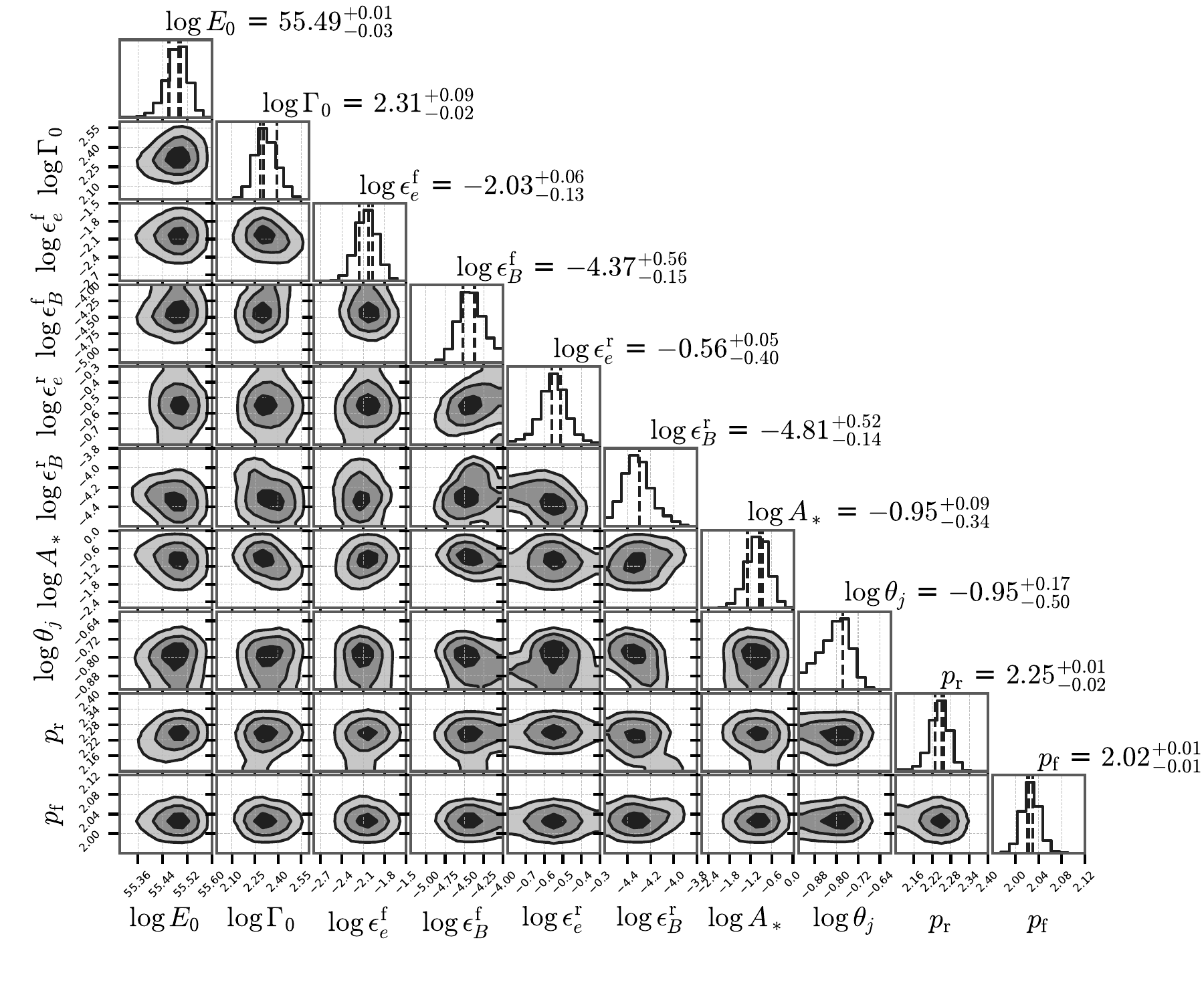}
     \caption{Corner plot of the MCMC posterior distributions for the forward--reverse shock afterglow model of GRB~240825A.}
     \label{Fig4:Corner_plot}
\end{figure}

To infer the jet properties of GRB~240825A, we model the multi-wavelength afterglow light curves within the standard external forward-reverse shock (FS-RS) framework \citep{1998ApJ...497L..17S,1999MNRAS.309..513H,2000ApJ...543...66P, 2001ApJ...548..787S,2009ApJ...703..675N, 2010ApJ...712.1232W,2013ApJ...776..120Y}. We adopt a wind-like circumburst medium with $n\propto r^{-2}$, assume a Milky Way-like host extinction curve, and include the empirical evolution of $E_{B-V}^{\rm host}$ inferred from the SED analysis. The model parameters include the initial bulk Lorentz factor (\(\Gamma_0\)), isotropic-equivalent kinetic energy (\(E_{\rm k,iso}\)), circumburst density parameter (\(A_*\)), electron and magnetic equipartition factors ($\epsilon_e$) and ($\epsilon_B$), and the power-law indices of the electron distributions  ($p$). Our fitting results are shown in the upper-left panel of Figure~\ref{Fig3:Afterglow}. The derived model parameters are reported in Table~\ref{tab:table3}, with their posterior distributions shown in Figure~\ref{Fig4:Corner_plot}. The broadband SEDs at three selected epochs are displayed in the bottom panels of Figure~\ref{Fig3:Afterglow}. The model well reproduces the observed afterglow emission from the radio to the GeV gamma-ray bands. The afterglow emission is initially dominated by the reverse shock and gradually becomes dominated by the forward shock at later times. The late optical emission at $t>2\times 10^{5}$ s exhibits a plateau-like behavior and is attributed to host-galaxy emission \citep{2025ApJ...979...38C}. Our model further predicts that sub-TeV gamma-ray afterglows at early times ($t< 10^3$ s) may be detectable with the current MAGIC and future CTA telescopes.     

\section{Late-time Optical Constraints on a Possible Supernova Component}
\label{sec:Late-time Optical Constraints on a Possible Supernova Component}
\begin{figure}[htbp]
     \centering
     \includegraphics[width=0.49\textwidth, angle=0]{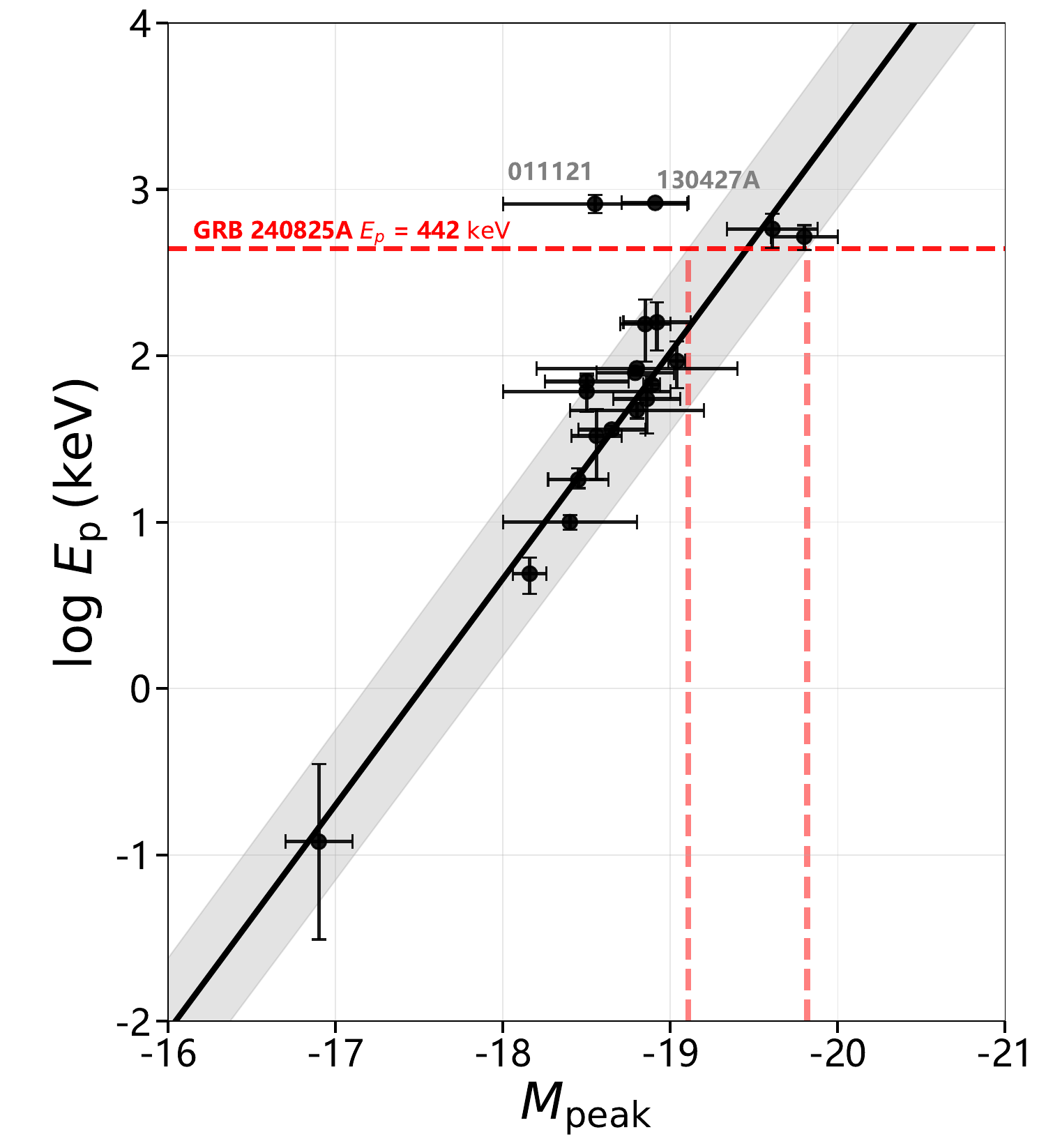}
     \includegraphics[width=0.49\textwidth, angle=0]{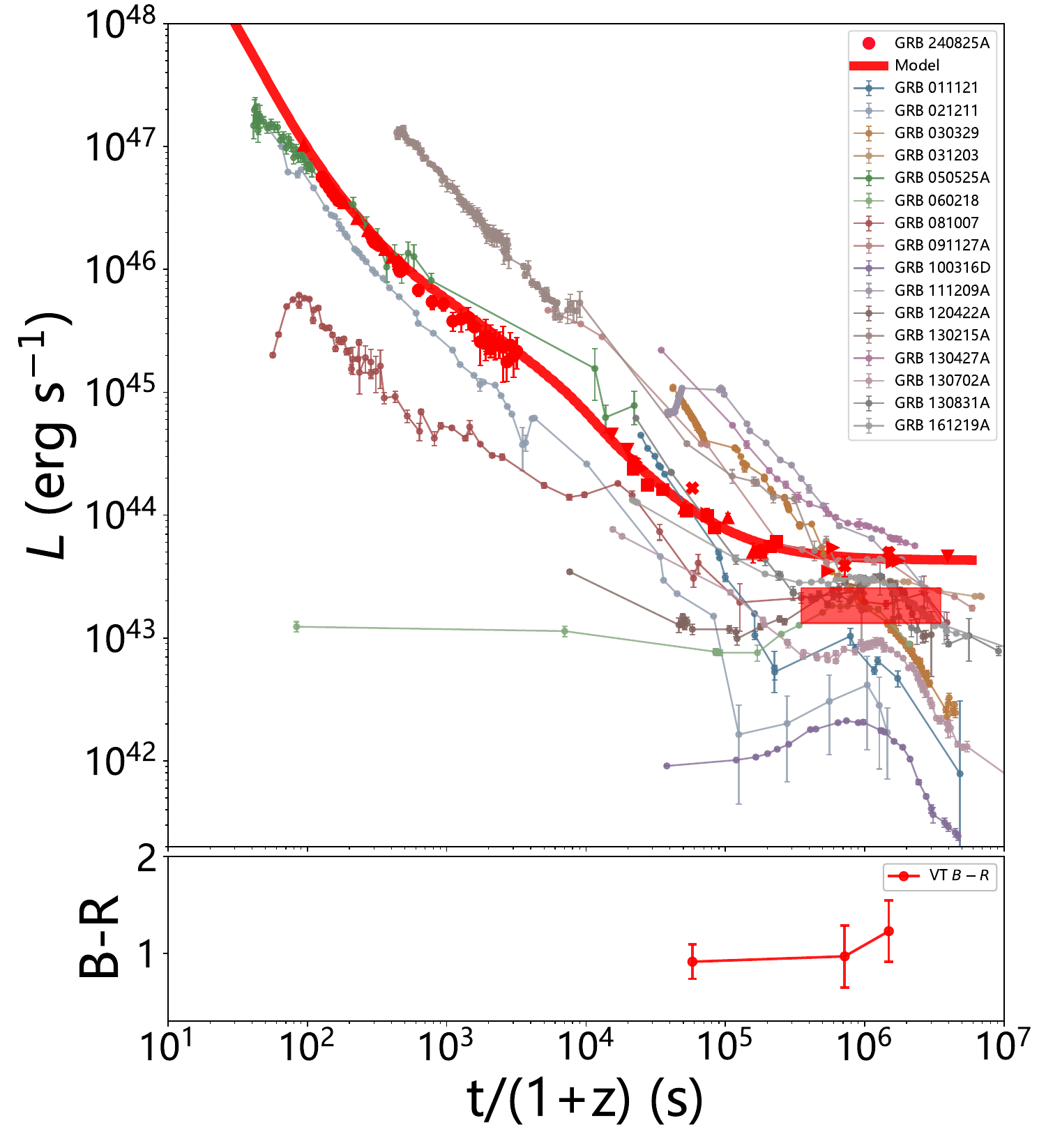}
     \caption{Left panel: $E_{\rm p}$--$M_{\rm peak}$ correlation. The solid line shows the best-fit power-law relation, while the shaded region represent the $2\sigma$ deviations. The red dashed line indicates the location of GRB~240825A ($E_{\rm p}=442$~keV). The vertical red dashed lines delimit the illustrative $M_{\rm peak}$ estimate and its uncertainty band derived from the empirical relation. Right panel: Rest-frame luminosity light curve of GRB~240825A compared with GRB/SN systems from the sample of \cite{2018ApJ...862..130L} (upper subpanel, the red shaded region indicates the inferred $M_{\rm peak}$ and its uncertainty band), and the color index evolution of the late-time VT data (lower subpanel).}
     \label{Fig5:SN}
 \end{figure}

GRB~240825A resides in a massive, dusty, star-forming galaxy with properties typical of collapsar GRB hosts, yet no clear supernova signature has been detected\citep{2025arXiv250800142G}. Our broadband afterglow analysis constrains the circumburst medium and the line-of-sight dust extinction, supporting a wind density profile and indicating time-dependent dust extinction. We therefore focus on its late-time optical emission to search for a possible supernova contribution and assess how dust extinction and host-galaxy light affect its detectability. An empirical relation between the prompt-emission peak energy ($E_{\rm p}$) and the peak bolometric absolute magnitude ($M_{\rm peak}$) of GRB-associated supernovae has been reported \citep{2006MNRAS.372.1357L,2018ApJ...862..130L}. We use the relation of \citet{2018ApJ...862..130L} for an illustrative estimate, assuming that a putative SN associated with GRB~240825A is consistent with the calibration sample.\footnote{The relation is based on a limited calibration sample of GRB-associated SNe and may be affected by observational selection, since bright, weakly extinguished events with little host contamination are easier to characterize. Its applicability to faint or heavily obscured SNe is therefore uncertain.} For $E_{\rm p}=442\pm8$~keV, the relation gives a nominal peak magnitude of $M_{\rm peak}=-19.46\pm0.35$~mag for a putative SN associated with GRB~240825A, corresponding to $L_{\rm SN,peak}=1.84_{-0.51}^{+0.71}\times10^{43}\ {\rm erg\,s^{-1}}$. This value is marked in the left panel of Figure~\ref{Fig5:SN}. The right panel compares the rest-frame $R$-band luminosity evolution of GRB~240825A with the GRB-SN sample compiled by \citet{2018ApJ...862..130L}. The red shaded region indicates the nominal $M_{\rm peak}$ range inferred from the empirical relation. At this estimated luminosity, the putative SN would be about five times fainter than the host galaxy's optical emission, making it difficult to distinguish from the host light. Host-galaxy extinction would further reduce the observed SN flux. The lack of a clear late-time SN bump therefore does not, by itself, rule out an underlying host-obscured supernova.

We present the $B-R$ color evolution derived from the deep SVOM/VT observations in the right panel of Figure~\ref{Fig5:SN}. The color index remains nearly stable around $\sim 0.97$ over $\sim 7.20 \times 10^5-1.49 \times 10^6$ s, and shows a weak increase to $1.21\pm 0.31$ at later epochs. The present optical observations do not establish an associated SN, but they also do not exclude a component that is faint or difficult to separate from the host emission. 

\section{Discussion and Conclusions}
\label{sect:Discussion and Conclusions}
Our temporal and hardness-ratio analysis of prompt emission in GRB~240825A supports a two-stage central-engine evolution, which can be interpreted as an initial proto-magnetar phase evolving into a fallback-accretion-powered BH central engine. With the reference configuration $(M_{\rm BH,0},a_0)=(3.0\,M_\odot,0.53)$ based on the expectations of \citet{2006ApJ...637..914W}, our fallback-fed disk model reproduces the main light-curve profile of the later episode. Flares superposed on the declining emission may arise from variations in the accretion flow \citep{2006ApJ...636L..29P,2006MNRAS.370L..61P} or intermittent magnetic dissipation within the jet \citep{2011ApJ...726...90Z}. The inferred fallback properties depend on the adopted initial BH conditions, particularly the spin. \citet{2023ApJ...952L..32G} showed that once the magnetic flux threading the BH reaches saturation in a magnetically arrested disk, the jet power is governed by the BH spin and mass accretion rate. In our fits using the adopted BZ prescription, lower initial spins generally require higher peak inner accretion rates to reproduce the observed luminosity. This is qualitatively consistent with the results of \citet{2024MNRAS.527.7905L}, who found that BH mass growth is largely independent of the initial BH mass but is greater for lower initial spins.

The afterglow analysis requires a time-dependent host extinction, $E^{\rm host}_{B-V}(t)$, to reproduce the broadband emission in a wind-like medium. The inferred color excess rises from $0.41\pm0.01$ to $0.63\pm0.02$ and then falls to $0.37\pm0.08$. This non-monotonic evolution is inconsistent with a static sightline and, together with the reported variations of $\beta_{\rm OX}$ and $N_{\rm H}^{\rm host}$ \citep{2026ApJ...997..101L}, favors a dynamically evolving circumburst environment. The mid-interval peak may be understood within the radiative-torque disruption (RATD; \citealt{2019NatAs...3..766H}) framework. Intense afterglow radiation fragments large grains into smaller ones while approximately conserving the total dust mass. During the initial disruption phase, the enhanced abundance of small grains can temporarily raise $A_B$ relative to $A_V$, producing a transient increase in $E(B-V)$ before both bands decline as disruption proceeds \citep{2020ApJ...895...16H}. The subsequent reddening decline is consistent with continued dust destruction along the line of sight \citep{2000ApJ...537..796W, 2014MNRAS.440.1810M, 2026ApJ...997..101L}. The red-to-blue color evolution observed in SN~2023ixf was interpreted as evidence for shock interaction with a pre-existing dust shell \citep{2024Natur.627..754L}. A similar episodic mass-loss history could contribute to this inhomogeneous environment \citep{2013ApJ...766...30G, 2013ApJ...778...67N}, but the observations do not uniquely distinguish circumstellar structures from larger-scale structures in the host galaxy.

No distinct supernova signature is identified in the late-time optical observations of GRB~240825A, which are dominated by host-galaxy emission. The current data therefore cannot distinguish between a host-obscured normal supernova and a failed-explosion scenario \citep{2025arXiv250800142G}. The latter remains an intriguing possibility, as a failed explosion would leave a significant fraction of the stellar envelope gravitationally bound, enabling its fallback onto the newly formed black hole disk and providing the energy reservoir needed to power the inferred relativistic jet. Recent observations of the disappearing massive star M31-2014-DS1 further demonstrate that massive stars can collapse into black holes without producing luminous supernovae \citep{2026Sci...391..689D}. At present, both a host-obscured normal supernova and a failed-explosion/fallback scenario remain viable interpretations of the late-time observations, and both are compatible with the fallback-accretion scenario inferred from our prompt-emission modeling.

In summary, GRB~240825A provides a valuable case for investigating central-engine activity and the circumburst environment of long GRBs. Its distinct prompt-emission episodes motivate a possible transition from a transient proto-magnetar to a BH engine. Under the adopted assumptions, fallback accretion can reproduce the main light-curve profile of the later episode. The observed evolving extinction further suggests a structured, dusty circumburst environment. 

No distinct supernova component is identified from the observations in the B and R bands with the SVOM/VT since bright host galaxy emission. Future deep and long-term optical monitoring with instruments such as SVOM/VT and \textit{JWST}/NIRSpec, particularly through late-time multi-band and near-infrared observations, will be crucial for distinguishing between a host-obscured supernova and a failed-explosion/fallback scenario. These observations provide new insights into the connection between massive-star evolution, BH formation, and relativistic jet production. We should note that while this manuscript was under review, \citet{2026arXiv260935697S} reported a dust-obscured broad-lined Type~Ic supernova associated with GRB~240825A from \textit{JWST}/NIRSpec spectroscopy obtained 66.5 days after the burst (40.1 days in the rest frame). Their template-based analysis indicates an extinction-corrected SN luminosity comparable to that of SN~1998bw, broadly comparable in scale to our illustrative SN estimate. This detection supports a massive-star origin. Their analysis indicates that dust extinction and host-galaxy contamination help explain the earlier ground-based nondetections. 

\begin{acknowledgments}
This work is supported by the National Key R\&D Program (2024YFA1611704 and 2024YFA1611700), the Guangxi Science and Technology Innovation Platform Program (Leitai Action Plan, Grant No. Guike LT2600640026 ), Guangxi Key R\&D Program (Guangxi Funeng Action Plan, Grant No. FN2504240030), and National Natural Science Foundation of China (grant Nos. 12633009, 12133003, 12203015). E. W. L. is also supported by the Guangxi Talent Program (``Highland of Innovation Talents'').
\end{acknowledgments}

\begin{table}[htbp]
     \footnotesize
     \centering
     \caption{Best-fit parameters of the power-law model for the afterglow of GRB~240825A.
     The columns list the temporal decay indices of the afterglow (\(\alpha_{\rm X,i}\) and \(\alpha_{\rm O,i}\)),
     the spectral index (\(\Gamma\), or equivalently \(\beta\)),
     the host hydrogen column density (\(N^{\rm host}_{\rm H}\)),
     and the color excess ($E^{\rm host}_{\rm B-V}$).}
     \begin{tabular}{cccccc}
     \toprule
     Time Interval & Data & Decay index & $\Gamma$ ($\beta$) & $N^{\rm host}_{\rm H}$ ($10^{22}\,\mathrm{cm^{-2}}$) & $E_{B-V}^{\rm host}$ \\
     \midrule
     $T_0 + [70, 200]~\mathrm{s}$ 
     & \makecell{X-ray (0.3--10~keV) \\ Optical (u, g, r, z)} 
     & \makecell{$\alpha_{\rm X,1} = 2.166^{+0.067}_{-0.057}$ \\ $\alpha_{\rm O,1} = 1.632^{+0.147}_{-0.146}$} 
     & $1.53\,(0.53)\pm0.01$ 
     & 0.353$\pm$0.016
     & 0.41 $\pm$0.01\\
     \hline
     $T_0 + [403, 10^{3}]~\mathrm{s}$ 
     & \makecell{X-ray (0.3--10~keV) \\ Optical (u, g, r, i, z)} 
     & \makecell{$\alpha_{\rm X,2} = 1.117^{+0.050}_{-0.050}$ \\ $\alpha_{\rm O,2} = 1.193^{+0.169}_{-0.172}$} 
     & $1.57\,(0.57)\pm0.01$ 
     & 0.642$\pm$0.015 
     & 0.63 $\pm$0.02\\
     \hline
     $T_0 + [4\times10^{3}, 2\times10^{4}]~\mathrm{s}$ 
     & \makecell{X-ray (0.3--10~keV) \\ Optical (g, r, i)} 
     & \makecell{$\alpha_{\rm X,3} = 1.263^{+0.070}_{-0.071}$ \\ $\alpha_{\rm O,3} = 1.081^{+0.209}_{-0.232}$} 
     & $1.70\,(0.70)\pm0.06$ 
     & 0.558$\pm$0.069
     & 0.37 $\pm$0.08\\
     \bottomrule
     \end{tabular}
     \label{tab:table1}
     \medskip
     \parbox{\linewidth}{\footnotesize\noindent\textit{Note.}
     Temporal decay indices are obtained from MCMC fitting of the optical and XRT afterglow data. Independent fits of different optical bands show no significant differences in the temporal decay indices within the same intervals, although with larger uncertainties. We therefore average the optical fluxes within each interval and perform joint optical--XRT spectral fitting with the \texttt{Xspec}, yielding the photon index $\Gamma$, host hydrogen column density \(N^{\rm host}_{\rm H}\), and host-galaxy color excess $E^{\rm host}_{\rm B-V}$.}
\end{table}

\begin{table}[htbp]
\centering
\caption{Physically motivated initial BH parameters and corresponding BZ fallback-accretion fits.}
\label{tab:table2}
\begin{tabular}{lccccccc}
\toprule
Model/source 
& $M_{\rm BH,0}$($M_\odot$)  
& $a_0$ 
& $M_{\rm fb}$($M_\odot$) 
& $r_{\rm fb}$($r_g$)  
& $t_{\rm fb}$(s)
& $s$ 
& $\dot{M}_{\rm in}^{\rm peak}$($M_{\odot} \rm s^{-1}$)
\\
& 
&  
& [0.01,1.00]
& [8.0,30.0]
& [0.01,2.00] 
& [0.0,1.0] \\
\midrule
35OC-RO  & 3.19 & 0.48 & $0.542^{+0.044}_{-0.188}$ & $16.964^{+0.277}_{-1.568}$ & $0.231^{+0.005}_{-0.012}$ & $0.405^{+0.024}_{-0.104}$ & $2.26 \times 10^{-1}$ \\
35OC-RO2 & 2.90 & 0.41 & $0.592^{+0.006}_{-0.115}$ & $17.562^{+0.347}_{-0.371}$ & $0.193^{+0.003}_{-0.006}$ & $0.347^{+0.004}_{-0.026}$ & $3.19 \times 10^{-1}$ \\
35OC-Sw  & 2.38 & 0.13 & $0.822^{+0.169}_{-0.521}$ & $8.236^{+2.610}_{-0.185} $ & $0.097^{+0.005}_{-0.033}$ & $0.366^{+0.055}_{-0.276}$ & $1.38 \times 10^{\rm0}$ \\
35OB-RO  & 2.49 & 0.26 & $0.960^{+0.036}_{-0.458}$ & $16.980^{+0.337}_{-1.533}$ & $0.126^{+0.003}_{-0.013}$ & $0.385^{+0.005}_{-0.119}$ & $6.92 \times 10^{-1}$ \\
35OC-RO, WH06 & 3.00 & 0.53 & $0.454^{+0.046}_{-0.097}$ & $17.565^{+0.413}_{-2.351}$ & $0.245^{+0.006}_{-0.006}$ & $0.409^{+0.034}_{-0.039}$ & $1.77 \times 10^{-1}$ \\
\bottomrule
\end{tabular}
\vspace{0.5em}
\begin{minipage}{0.98\linewidth}
\footnotesize
\raggedright
\textit{Note.}
The first four parameter sets are taken from the BH-forming models of \cite{2021MNRAS.500.4365A}, with the PNS mass and dimensionless angular momentum at collapse adopted as proxies for the initial BH mass \(M_{\rm BH,0}\) and spin \(a_0\). The last row gives the expectations of \cite{2006ApJ...637..914W} for model 35OC-RO, based on the mass and angular momentum enclosed within the inner \(\sim3\,M_\odot\) of the pre-supernova progenitor. For each \((M_{\rm BH,0},a_0)\), the fallback parameters are refitted to the observed prompt emission, starting from \((M_{\rm fb},r_{\rm fb},t_{\rm fb},s)=(0.1\,M_\odot,12.0\,r_g,0.15\,{\rm s},0.5)\); the allowed ranges are given below the corresponding columns. The Gaussian-like pulse components are excluded from the fallback-parameter search. Here, $M_{\rm fb}$ is the total mass supplied to the disk during the modeled post-collapse phase, $r_{\rm fb}$ is the circularization radius at which fallback material enters the disk, $t_{\rm fb}$ is the characteristic mass-supply timescale, and $s$ is the disk-wind index. The final column gives the peak inner accretion rate, $\dot{M}_{\rm in}^{\rm peak}$. Model 35OC-Sw (\(a_0=0.13\)) is included only as a slowly rotating limiting case.
\end{minipage}
\end{table}

\begin{table}[htbp]
     \footnotesize
     \centering
     \caption{Multi-band Fitting Results with the Forward--Reverse Shock Model}
     \begin{tabular}{lc}
     \toprule
     Parameter & Posterior value \\
     \midrule
     $\log E_0$               & $55.49^{+0.01}_{-0.03}$ \\
     $\log \Gamma_0$          & $2.31^{+0.09}_{-0.02}$ \\
     $\log \epsilon_e^{\rm f}$ & $-2.03^{+0.06}_{-0.13}$ \\
     $\log \epsilon_B^{\rm f}$  & $-4.37^{+0.56}_{-0.15}$ \\
     $\log \epsilon_e^{\rm r}$ & $-0.56^{+0.05}_{-0.40}$ \\
     $\log \epsilon_B^{\rm r}$  & $-4.81^{+0.52}_{-0.14}$ \\
     $\log A_*$               & $-0.95^{+0.09}_{-0.34}$ \\
     $\log \theta_j$          & $-0.95^{+0.17}_{-0.50}$ \\
     $p^{\rm r}$              & $2.25^{+0.01}_{-0.02}$ \\
     $p^{\rm f}$              & $2.02^{+0.01}_{-0.01}$ \\
     \bottomrule
     \end{tabular}
     \label{tab:table3}
     \parbox{\linewidth}{\footnotesize\noindent\textit{Note.} The superscripts ``f'' and ``r'' denote the forward-shock and reverse-shock components, respectively.}
\end{table}

\clearpage

\bibliography{bibtex}
      
\end{document}